\documentclass[aps,pra,showpacs,superscriptaddress,groupedaddress]{revtex4}  
\usepackage{graphicx}  
\usepackage{dcolumn}   
\usepackage{bm}        
\usepackage{amssymb}   
\usepackage{color}
\usepackage{braket}
\usepackage{amsmath}
\usepackage[caption=false]{subfig}

\newcommand*\diff{\mathop{}\!\mathrm{d}}
\newcommand*\Diff[1]{\mathop{}\!\mathrm{d^#1}}
\newcommand{\ketbra}[2]{| #1 \rangle \langle #2 |}
\newcommand{\alketbra}[2]{| \alpha_{#1} \rangle \langle \alpha_{#2} |}

\DeclareMathOperator{\Tr}{Tr}
\DeclareMathOperator{\sech}{sech}

\begin{document}


\title{Coarse-graining of measurement and quantum-to-classical transition in the bipartite scenario}

\author{Madhav Krishnan V} 
\email{madhavkv@imsc.res.in}
\affiliation{Optics and Quantum Information Group, The Institute of Mathematical Sciences, HBNI, Taramani, Chennai 600113, India}

\author{Tanmoy Biswas} 
\email{tb12ms052@iiserkol.ac.in}
\affiliation{ Department of Physical Sciences, Indian Institute of Science Education and Research Kolkata, Mohanpur - 741 246, WB, India}

\author{Sibasish Ghosh} 
\email{sibasish@imsc.res.in}
\affiliation{Optics and Quantum Information Group, The Institute of Mathematical Sciences, HBNI, Taramani, Chennai 600113, India}
\date{\today}

\begin{abstract}
The connection between coarse-graining of measurement and emergence of classicality has been investigated for some time, if not well understood. Recently in (PRL \textbf{112}, 010402, (2014)) it was pointed out that coarse-graining measurements can lead to non-violation of Bell-type inequalities by a state which would violate it under sharp measurements. We study here the effects of coarse-grained measurements on bipartite cat states. We show that while it is true that coarse-graining does indeed lead to non-violation of a Bell-type inequality, this is not reflected at the state level. Under such measurements the post-measurement states can be non-classical (in the quantum optical sense) and in certain cases coarse-graning can lead to an increase in this non-classicality with respect to the coarse-graining parameter. While there is no universal way to quantify non-classicality, we do so using well understood notions in quantum optics such as the negativity of the Wigner function and the singular nature of the Gluaber-Sudharshan P distribution. 
\end{abstract}

\pacs{03.65.Ud, 03.65.Ta, 03.67.Mn, 42.50.−p, 42.50.Dv}
\maketitle

\section{Introduction}
Quantum mechanics has been perhaps the most successful theory of nature that modern science has developed and it is believed to describe the universe from atoms to galaxies. But from its advent, there has been a puzzling question raised about the theory; why is it that the full richness of quantum mechanics only expresses itself in the microscopic world while classical mechanics is sufficient to describe the macroscopic world? Is there some boundary between the purely quantum regime and the regime where classical mechanics is valid? If so, how does the transition between these two regimes take place? These questions are of both fundamental theoretical interest, as well as of practical interest for the development of quantum technologies which need to be robust against this classical transition. In the early days of quantum mechanics, the lack of macroscopic quantum effects were attributed to the smallness of the Planck's constant and mathematically the classical limit was seen as $\hbar \rightarrow 0$. A statistical correspondence between expectation values of quantum observables and their macroscopic counterparts was shown by Ehernfest under the assumption of large eigenvalues. However, these arguments could not rule out macroscopic superpositions, as pointed out by Schr\"{o}dinger with his famous thought experiment involving a superposition of dead and alive states of a cat. Decoherence theory \cite{zurek} offered an answer, based on dynamics which would lead to the suppression of such macroscopic superpositions. A complimentary line of thought that developed was to attribute the classical transition to the limited precision or coarse-graining of most macroscopic measurements. This is the approach we will be examining in this paper.

While superpositions and their corresponding interference effects play a great role in non-classical effects, non-classicality is not limited to interference phenomenon. Bell's famous discovery, that a local realistic description of two spin-$\frac{1}{2}$ system was inconsistent with the predictions of quantum mechanics \cite{bell_original}, identified non-locality as an important feature of non-classicaliy. Mermin extended this result for arbitrary spin values J and showed that even under the so called classical limit of J $\rightarrow \infty$, the spins will violate a generalized Bell-type inequality \cite{mermin_1980}. However there was an important caveat; the measurement precision required to see this violation was also proportional to J, hence without high enough precision the `classicality' of the macroscopic world was preserved. Following a similar line of thought and using macrorealism as a requirement for a classical description, Kofler et al. \cite{kofler} showed that increasing the system dimension does not necessarily lead to a classical limit in terms of violation of Leggett-Garg inequalities by temporal correlations of consecutive measurements on a spin-J system. However, they showed that using imprecise or coarse-grained measurements instead of their sharp counterparts can give rise to a classical description. More recently, Jeong et al. \cite{kim} proposed that coarse-graining of measurement can be performed in two distinct manners:

 1) \textit{Coarse-graining of the resolution of measurement}, where the measuring device cannot distinguish between elements of the measurement basis with complete accuracy. For example, not being able to distinguish between different closely spaced eigenstates of a spin-$z$ measurement is an instance of resolution coarse-graining. 2) \textit{Coarse-graining of reference of measurement} where the experimenter does not have complete control over the basis that measurement is being carried out in. An example would be again a spin-$z$ measurement but here, there could be an uncertainty of a small solid angle $d \Omega$ about the $z$ axis. It was shown in \cite{kim} that under these types of coarse-grained measurements, a state which would violate a Bell-type inequality under sharp measurements will no longer do so as the coarse-graining parameter is increased. This was argued to be a signature of quantum-to-classical transition. However, while it is true that if a state violates a Bell-type inequality, it can be called a non-classical state in the sense of not allowing a local realistic description, the converse is not true in general. A state that does not violate a Bell-type inequality cannot be called a classical state for two reasons. First, to rule out local realistic models one has to check, with all possible independent Bell-type inequalities - whose number grows exponentially with system dimension. Second, there exist non-classical states which do not violate any Bell-type inequalities, for example Werner states which are entangled (and thereby, non-classical) but allows a local realistic description \cite{werner_1989}.

 In this work we consider bipartite cat states which violate a CHSH-type Bell's inequality under sharp measurement, with the property that the violation disappears under coarse-graining of the measurement. We calculate the post-measurement state under both reference and resolution coarse-graining of the measurement. We show these states to be non-classical by calculating the Gluaber-Sudharshan P-distribution and the Wigner functions of these states and quantify this non-classicality using the negative phase space volume of the Wigner function. Our results show that depending on the choice of measurement operator, the non-classicality of the post-measurement state can be made to increase under coarse-graining contrary to what the non-violation of the Bell-type inequality suggests. Apart from considering the bipartite cat state which is an equal superposition of the two tensor products of the even and odd coherent states (by interchanging the orders in the tensor product), we also consider here a general version of this cat state as well as a cat state which is equal superposition of tensor products of the single photon-added and two photons-added squeezed vacuum states (see sections \ref{ent_fock_sec} and \ref{squeezed_sec}). We find that non-classicality of the post-measurement state increases with higher photon number in the second case but curiously, it decreases with higher values of the squeezing parameter in the last case. 

\subsection{Coarse-Graining}
Consider an infinite dimensional Hilbert space with an orthonormal basis $\left\lbrace | o_n \rangle \right\rbrace$ with $ n \in \mathbb{Z} $, the set of all integers. A dichotomous measurement operator in this Hilbert space can be constructed as
\begin{equation}
\begin{split}
O^k &= O_{+}^{k} - O_{-}^{k} ,  \\
\text{where,   } \hspace{0.5cm} O_{+}^{k}&=  \sum\limits_{n = k}^{+ \infty } |o_n\rangle\langle o_n |  \hspace{0.5cm} ; \hspace{0.5cm} O_{-}^{k} = \sum\limits_{n = -\infty}^{k - 1} |o_n\rangle\langle o_n |.
\end{split}
\end{equation}
The resolution coarse-grained measurement operator can be written as 
\begin{equation}\label{odelta_def}
O_{\delta} = \sum\limits_{k = -\infty}^{\infty} P_{\delta}(k) O^{k},
\end{equation}
where, $P_{\delta}(k) = N_{\delta} \exp \left( -\frac{k^2}{2\delta} \right) $ is a normalized discrete Gaussian distribution with variance $\delta^{2}$. The reference coarse-grained measurement operator will be
\begin{equation}\label{oDelta_def}
O_{\Delta}(\theta_a) = \int P_{\Delta}(\theta - \theta_a) \left[ U^{\dagger}(\theta) O^0 U(\theta)\right],
\end{equation}
where $P_{\Delta}(\theta - \theta_a)$ is a Guassian distribution centered about $\theta_a$ with variance $\Delta^2$ and $U(\theta)$ is a unitary rotation whose effect on the basis states will be
\begin{equation}
\begin{split}
&U(\theta) \ket{o_n} = \cos \theta \ket{o_n} + \sin\theta \ket{o_{-n}}, \\
&U(\theta) \ket{o_{-n}} = \sin \theta \ket{o_n} - \cos\theta \ket{o_{-n}} .
\end{split}
\end{equation}
The paper is structured in the following manner. In section \ref{bell_sec}, we calculate violation of a Bell-type inequality by the bipartite cat states formed out of even and odd coherent states, under coarse-grained measurement, and show  their non-violation with increasing coarse-graining. In section \ref{fock_sec}, the post-measurement state for a single-mode Fock state is calculated and shown to be non-classical. In section \ref{post_cat_sec}, the post-measurement states for bipartite cat states under both resolution and reference coarse-graining is calculated and we find their Glauber-Sudharshan P-distributions as well as their Wigner functions. Further, the relation between coarse-graining and the negativity of the Wigner function is examined. In section \ref{ent_fock_sec}, the initial state is taken to be a NOON state and it is shown that the post-measurement state non-classicality increases with initial state photon number. In  section \ref{squeezed_sec} a similar analysis shows that post-measurement state non-classicality is decreased by increased squeezing in the initial state, the later being chosen to be entangled photon added squeezed vacuums. Finally in section \ref{conclusions}, a summary of our results and conclusions are given with some future directions we will be exploring.
\section{Bell-type inequality non-violation by Schr\"{o}dinger cat states} \label{bell_sec}

\begin{figure}[!tbp]
  \centering
  \begin{minipage}[b]{0.48\textwidth}
    \includegraphics[width=\textwidth]{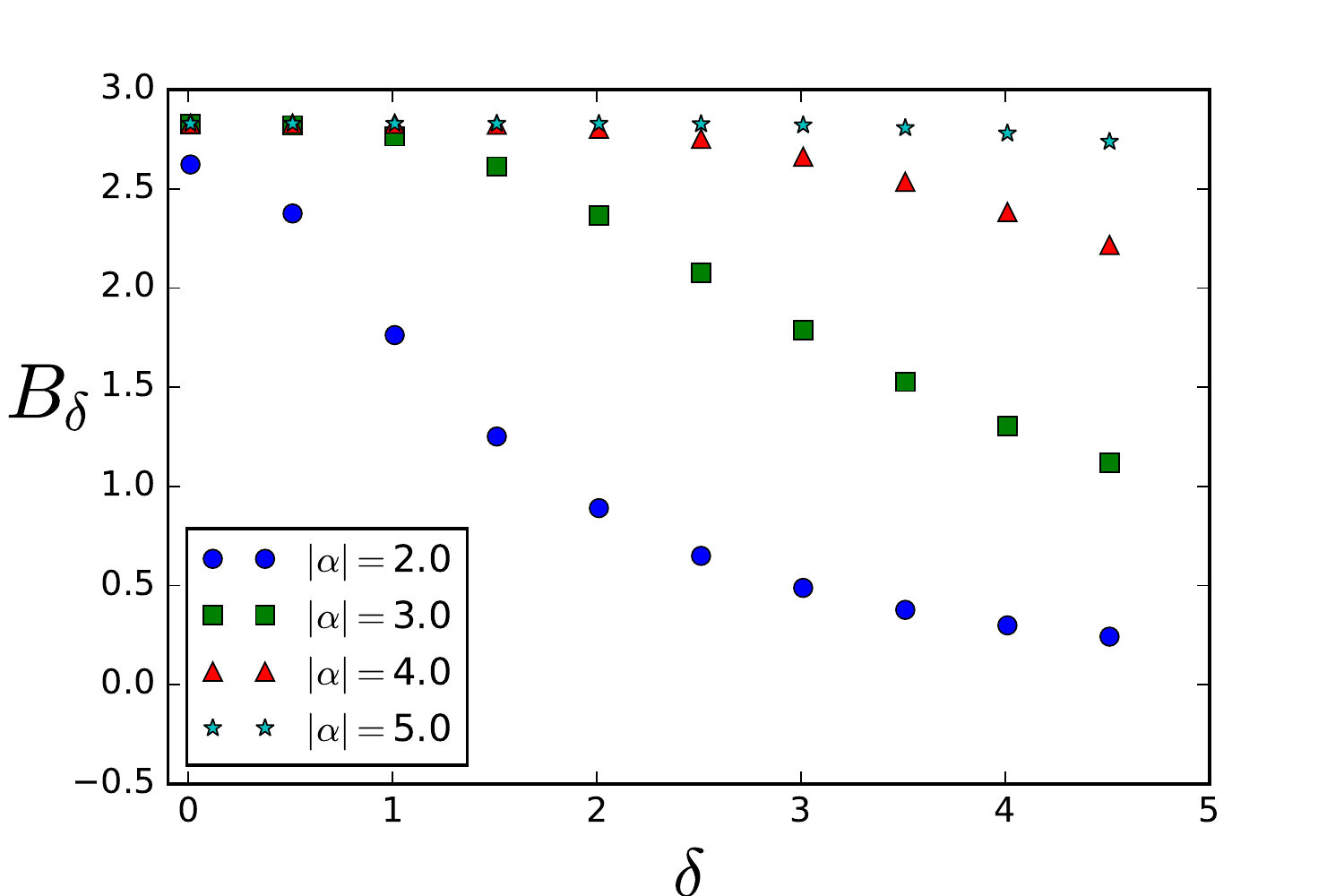}
    \caption{Resolution coarse-graining: plotting $B_{\delta}$ of equation \eqref{bell_delta_eqn} with respect to $\delta$ (color online). }
    \label{bell_delta_fig}
  \end{minipage} 
  \hfill
  \begin{minipage}[b]{0.48\textwidth}
    \includegraphics[width=\textwidth]{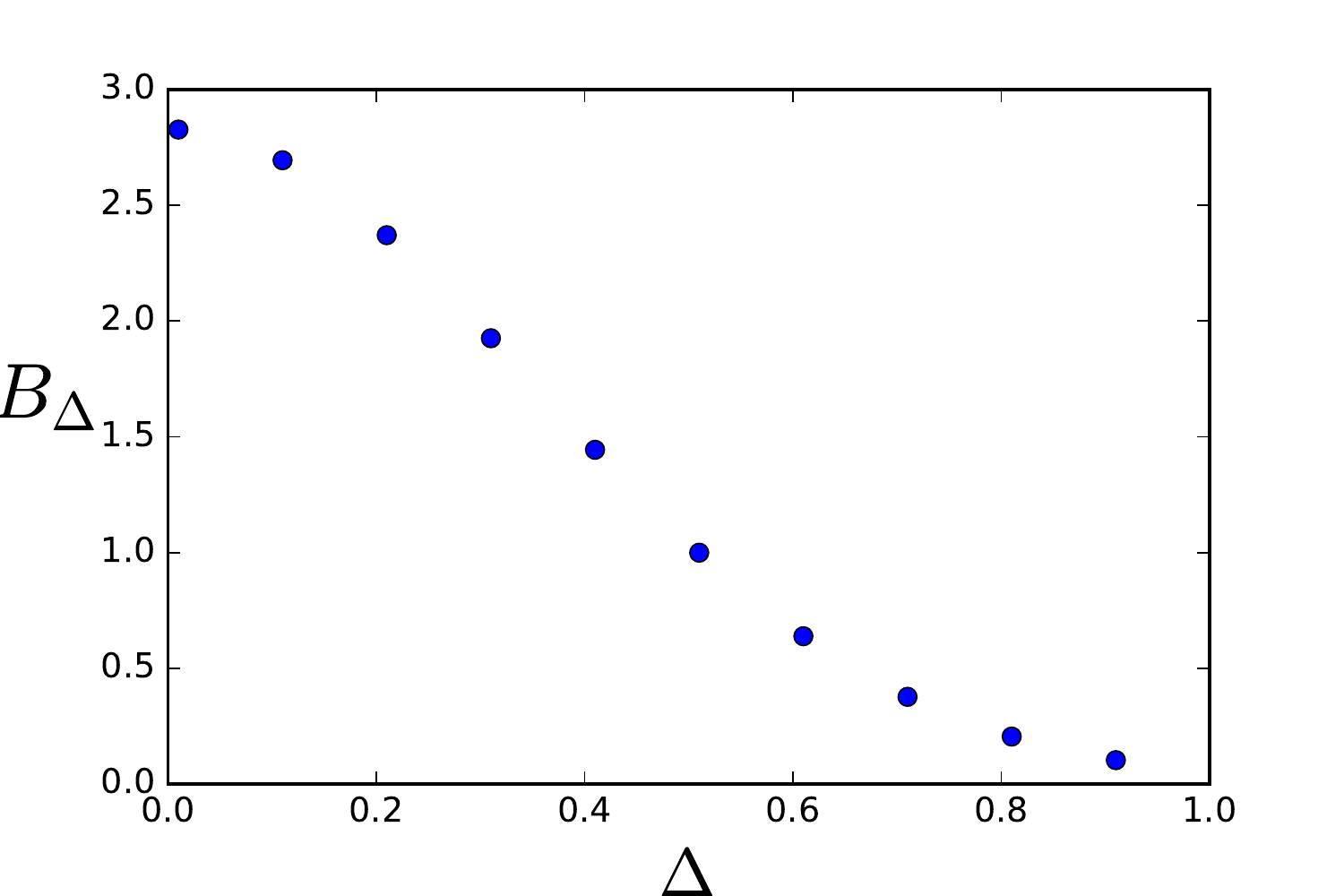}
    \caption{Reference coarse-graining: plotting $B_{\Delta}$ of equation \eqref{bell_Delta_eqn} with respect to $\Delta$.}
    \label{bell_Delta_fig}
  \end{minipage}
\end{figure}

Cat states or even and odd coherent states of a single mode system are defined as
\begin{equation}
\label{cat_state_def_eqn}
\begin{split}
\ket{\alpha_{e}} = N_e \left( \ket{\alpha} + \ket{-\alpha} \right), \\
\ket{\alpha_{o}} = N_o \left( \ket{\alpha} - \ket{-\alpha} \right), 
\end{split}
\end{equation}
where, in terms of Fock states, 
\begin{equation}\label{coherent_state_def}
\ket{\alpha} = e^{-\frac{|\alpha |^2 }{2}}\sum\limits_{n = 0}^{+\infty}\frac{\alpha^n}{\sqrt{n!}}\ket{n}, 
\end{equation}
is the coherent state of the system corresponding to the scalar $\alpha \in \mathbb{C}$ and $N_e = \frac{1}{\sqrt{2 + 2 e^{-2|\alpha|^2}}}$ and $N_o = \frac{1}{\sqrt{2 - 2 e^{-2|\alpha|^2}}} $ are normalization factors. Note that we have taken our basis to be even and odd Fock states i.e., $\ket{o_n} = \ket{2n}$ for $n = 0, 1, 2,...$  and $\ket{o_{-n}} = \ket{2n +1 }$ for $n = 1, 2, 3,...$ Consider the following bipartite cat state, 
\begin{equation}\label{psi}
\ket{\psi_{\alpha}} = \frac{1}{\sqrt{2}} \left( \ket{\alpha_e}\ket{\alpha_o} + \ket{\alpha_o}\ket{\alpha_e} \right).
 \end{equation} 
We can compute the Bell quantity in this state to be
\begin{equation}\label{bell_quanity_def}
B  = E_{ab} + E_{cb} + E_{ad} - E_{cd},
\end{equation} 
where, $ E_{ab} = \langle O^0(\theta_a) \otimes O^0(\theta_b) \rangle$, the expectation being taken with respect to $\ket{\psi_{\alpha}}$ and $O^{0}(\theta_i) = U^{\dagger}(\theta_i) O^{0} U(\theta_i)$. By using resolution and reference coarse-grained versions of these measurement operators and maximizing over $\theta_a, \theta_b, \theta_c, \theta_d$, we can obtain the maximum Bell violations $B_\delta$ and $B_\Delta$ for resolution and reference coarse-graining, respectively. We choose the unitary operator $U(\theta)$ to be a rotation  between even and odd coherent states, such that,

\begin{equation}\label{rot_alpha_def}
\begin{split}
&U(\theta) \ket{\alpha_e} = \cos \theta \ket{\alpha_e} + \sin\theta \ket{\alpha_o}, \\
&U(\theta) \ket{\alpha_o} = \sin \theta \ket{\alpha_e} - \cos\theta \ket{\alpha_o} .
\end{split}
\end{equation}
We find that (see appendix \ref{bell_appendix})
\begin{equation}\label{bell_delta_eqn}
\begin{split}
  B_{\delta} = \max\limits_{\theta_a, \theta_b, \theta_c, \theta_d} \mathcal{F}(\theta_a, \theta_b, \theta_c, \theta_d) \left(-1 + A + B - \frac{1}{4} (A + B)^2 \right)  + \frac{1}{2}(A - B)^2,
\end{split}
\end{equation}  
where
\begin{equation}
\begin{split}
A  &= 2C_e^2\sum\limits_{k = 0}^{\infty} \sum\limits_{n = 0}^{k} P_{\delta}(k)\frac{(|\alpha|^2)^{2n}}{(2n)!} ,\\
B  &= 2C_o^2 \sum\limits_{k = 1}^{ \infty} \sum\limits_{n = 0}^{k - 1} P_{\delta}(k)\frac{(|\alpha|^2)^{2n + 1}}{(2n + 1)!} ,\\
\end{split}
\end{equation}
 $C_e^2 = (\cosh |\alpha|^2 )^{-1} $  ,  $C_o = (\sinh |\alpha|^2 )^{-1}$
and,
\begin{equation}
\begin{split}
\mathcal{F}(\theta_a, \theta_b, \theta_c, \theta_d) = \cos\left(2\theta_a + 2\theta_b \right) + \cos\left(2\theta_c + 2\theta_b \right)   + \cos\left(2\theta_a + 2\theta_d \right) - \cos\left(2\theta_c + 2\theta_d \right),
\end{split}
\end{equation}
with $\max\limits_{\theta_a, \theta_b, \theta_c, \theta_d} \mathcal{F}(\theta_a, \theta_b, \theta_c, \theta_d) = 2\sqrt{2}$ .
Similarly, for reference coarse-graining we find that (see appendix \ref{bell_appendix})
\begin{equation}\label{bell_Delta_eqn}
B_{\Delta} =  2\sqrt{2}e^{-4\Delta^2} .
\end{equation}
In FIG. \ref{bell_delta_fig}, we have plotted $B_{\delta}$ versus $\delta$ for different values of $\alpha$, the complex parameter of the even and odd coherent states. And similarly, we have plotted $B_{\Delta}$ versus $\Delta$ in FIG. 2.
\section{post-measurement state of single mode Fock state} \label{fock_sec}

As mentioned earlier, while a state violating a Bell-type inequality can be taken as a proof of non-classicality, classicality of the state does not follow immediately from non-violation. A relevant question to ask in this context is whether a coarse-grained measurement on a non-classical state will lead to a classical state. To check this, let us now calculate the post-measurement state under coarse-grained measurement for an even Fock state, i.e, choosing the initial state to be $\ket{\psi_{i}} = \ket{o_n} = \ket{2n}$. For non-selective measurement with the operator $O_{\delta}$ the post-measurement density matrix will be
\begin{equation}
\begin{split}
\rho_{res}(\delta) &=  \sum\limits_{k = -\infty}^{\infty} P_{\delta}(k) \left(  O_{+}^{k} | 2n\rangle\langle 2n | O_{+}^{k}  + O_{-}^{k} | 2n\rangle\langle 2n | O_{-}^{k}  \right), \\
&= | 2n \rangle\langle 2n |.
\end{split}
 \end{equation}
The action on the initial state by the measurement can be thought of as acting with the projective valued measurement (PVM) $\left\lbrace O_{+}^{k}, O_{-}^{k}  \right\rbrace$, with probability $P_{\delta}(k)$, corresponding to a non-selective measurement of the resolution coarse-grained operator in equation  (\ref{odelta_def}). Unsurprisingly, the measurement of $O_{\delta}$ does not change the input state $\ket{\psi_{i}}$, as it is an eigenstate of $O^{k}$ for all $k$. Looking at the P-distribution of post-measurement state
\begin{equation}
\mathcal{P}_{res}(\gamma) =  L_{2n}(-\nabla_{\gamma}^2) \delta^{(2)}(\beta),
\end{equation}
where $\gamma$ is the phase space variable, $(\nabla^2_{\gamma})^k = \dfrac{\partial^{2k}}{\partial\gamma^k \partial\gamma^{*k}}$, and $L_{n}(x)$ is the $n^{\text{th}}$ Laguerre polynomial. A P-distribution that is more singular than the delta function can have no classical analogue \cite{glaub_pdist}. So, from the perspective of quantum optical non-classicality, such a measurement does not \textit{always} lead to quantum-to-classical transition at the state level. What about reference coarse-graining? The post-measurement state in this case will be
\begin{equation}\label{ref_post_meas_eq}
\begin{split}
 \rho_{ref}(\Delta) = \int d\theta P_{\Delta}(\theta - \theta_a) \left(  O^{0}_{+}(\theta) |2n \rangle\langle 2n |O_{+}^{0}(\theta) \right.  
 \left. + O^{0}_{-}(\theta) |2n \rangle\langle 2n |O_{-}^{0}(\theta) \right) ,
 \end{split}
 \end{equation} 
where $O_{\pm}^{k}(\theta) = V^{\dagger}(\theta) O_{\pm}^{k} V(\theta)  $. Here, $V(\theta)$ is a unitary, acting in the two dimensional space spanned by the even and odd Fock states $\ket{2n}$  and $\ket{2n - 1}$ respectively, and whose action is given by,

\begin{equation}
\begin{split}
&V(\theta) \ket{2n} = \cos \theta \ket{2n} + \sin\theta \ket{2n + 1} ,\\
&V(\theta) \ket{2n + 1} = \sin \theta \ket{2n} - \cos\theta \ket{2n + 1} .
\end{split}
\end{equation}
Here again, the measurement can be thought of as acting with the PVM $\left\lbrace O_{+}^{0}(\theta), O_{-}^{0}(\theta) \right\rbrace$, with probability $P_{\Delta}(k)$ corresponding to a non-selective measurement of the operator in equation (\ref{oDelta_def}). The post-measurement state, $\rho_{ref}$ in equation (\ref{ref_post_meas_eq}), can then be rewritten as (see appendix \ref{fock_appendix})
\begin{equation}\label{ref_int_eq}
 \rho_{ref}(\Delta) = \int d\theta P_{\Delta}(\theta - \theta_a) \rho_{\theta},
\end{equation}
where,
\begin{equation}
\rho_{\theta} = \frac{1}{4}\left(
\begin{array}{cc} 
3 + \cos 4\theta & \sin 4\theta \\
\sin 4\theta &1 - \cos 4\theta
\end{array}
 \right),
\end{equation}
is expressed in the two dimensional basis, $\left\lbrace \ket{2n}, \ket{2n + 1} \right\rbrace$. Performing the integral in equation (\ref{ref_int_eq}), the post-measurement state can be written as
\begin{equation}\label{ref_coherence_eq}
\rho_{ref} = \frac{1}{4}\left(
\begin{array}{cc} 
3 + e^{-8\Delta^2}\cos 4\theta_a & e^{-8\Delta^2}\sin 4\theta_a \\
e^{-8\Delta^2}\sin 4\theta_a &1 - e^{-8\Delta^2}\cos 4\theta_a
\end{array}
 \right).
 \end{equation} 
Equation (\ref{ref_coherence_eq}) shows that increase in the reference coarse-graining parameter $\Delta$, reduces the coherence in the state $\rho_{ref}(\Delta)$ . The P-distribution for the state $\rho_{ref}(\Delta)$ is (see appendix \ref{fock_appendix})

\begin{equation}
\begin{split}
\mathcal{P}_{ref}(\gamma) = \frac{1}{4} \left\lbrace (3 + e^{-8\Delta^2}\cos 4\theta_a) L_{2n}(-\nabla_{\gamma}^2)  - e^{-8\Delta^2}\sin 4\theta_a \left ( \frac{\partial}{\partial\gamma} + \frac{\partial}{\partial\gamma^*} \right )M_{2n}(\nabla_{\gamma}^2) \right. \\ \left.
+ (1  - e^{-8\Delta^2}\cos 4\theta_a )L_{2n - 1}(-\nabla_{\gamma}^2) \right\rbrace \delta^{(2}(\gamma),
\end{split}
\end{equation}
where, 
\begin{equation}
M_{n}(x) \equiv \sum\limits_{m = 0}^{n} \frac{ \sqrt{n+1}}{(m + 1)!} {n  \choose m} x^m .
\end{equation}
The P-distribution is again seen to be more singular than a delta function. So, while reference coarse-graining seems to have a decoherence effect in killing off-diagonal terms, it cannot be claimed that it leads to a classical state.

\section{Measurement of two-mode odd-even coherent states} \label{post_cat_sec}
As non-violation of Bell-type inequality involves correlation in a bipartite system, it is relevant to see how coarse-graining affects the post-measurement state of such a system. We consider the following state of a two-mode system:
\begin{equation*}
\ket{\psi_{\alpha}} = \frac{1}{\sqrt{2}} \left( \ket{\alpha_e}\ket{\alpha_o} + \ket{\alpha_o}\ket{\alpha_e} \right).
 \end{equation*}
where, $\ket{\alpha_{e}} = \frac{1}{\sqrt{2 + 2\exp (-2 |\alpha|^2 )}} \left( \ket{\alpha} + \ket{-\alpha} \right)  $ and $\ket{\alpha_{o}} = \frac{1}{\sqrt{2 - 2\exp (-2 |\alpha|^2 )}} \left( \ket{\alpha} - \ket{-\alpha} \right)  $  , where $\ket{\alpha}$ is the coherent state defined in equation (\ref{coherent_state_def}).
\subsection{Reference coarse-graining}
Let us first consider a reference coarse-grained observable of the form 
\begin{equation}
\begin{split}\label{ref_twomode_meas_def}
O_{\Delta}(\theta_a, \theta_b) = \int\limits_{ - \infty}^{+ \infty} d\theta_1 \int\limits_{ - \infty}^{+ \infty} d\theta_2 P_{\Delta}(\theta_1 - \theta_a)P_{\Delta}(\theta_2 - \theta_b)   \left[ O^{0}(\theta_1) \otimes O^{0}(\theta_2) \right],
\end{split}
\end{equation}
where $O^{0}(\theta) = U^{\dagger}(\theta) O^0 U(\theta)$; the unitary operator $U(\theta)$ is as defined in equation (\ref{rot_alpha_def}). For non-selective measurement, the post-measurement state will be (see appendix \ref{post_meas_cat_appendix})
\begin{equation}\label{post_meas_ref_def}
\rho_{ref} = \int\limits_{ - \infty}^{+ \infty} d\theta_1 \int\limits_{ - \infty}^{+ \infty} d\theta_2 P_{\Delta}(\theta_1 - \theta_a)P_{\Delta}(\theta_2 - \theta_b) \rho(\theta_1, \theta_2),
\end{equation}
where
\begin{equation}
\begin{split}
\rho(\theta_1, \theta_2) = \frac{1}{2}\left\lbrace \sin^2(\theta_1 + \theta_2) |\alpha_e\rangle\langle \alpha_e |_{\theta_1} \otimes |\alpha_e\rangle\langle \alpha_e |_{\theta_2}  \right. \\ 
\left. + \sin^2 (\theta_1 + \theta_2) |\alpha_o\rangle\langle \alpha_o |_{\theta_1} \otimes |\alpha_o\rangle\langle \alpha_o |_{\theta_2}  \right. \\ \left. + \cos^2 (\theta_1 + \theta_2) |\alpha_e\rangle\langle \alpha_e |_{\theta_1} \otimes |\alpha_o\rangle\langle \alpha_o |_{\theta_2} \right. \\
\left. + \cos^2 (\theta_1 + \theta_2) |\alpha_o\rangle\langle \alpha_o |_{\theta_1} \otimes |\alpha_e\rangle\langle \alpha_e |_{\theta_2} \right\rbrace,
\end{split}
\end{equation}
and $| *\rangle\langle *|_{\theta} \equiv U^{\dagger}(\theta) |*\rangle\langle *| U(\theta)$. 
The density matrix expressed in the unrotated basis,\newline $\left\lbrace \ket{\alpha_e \alpha_e }, \ket{\alpha_e \alpha_o }, \ket{\alpha_o \alpha_e }, \ket{\alpha_o \alpha_o } \right\rbrace$, with $|{\alpha}_e\rangle$ and $|{\alpha}_o\rangle$ being given by equation \eqref{cat_state_def_eqn}, will be,
\begin{equation}\label{post}
\rho_{ref} = \left(
\begin{array}{cccc}
a &\phantom{-}b &\phantom{-}c &\phantom{-}d  \\
b &\frac{1}{2}-a &\phantom{-}d &-c \\
c &\phantom{-}d &\frac{1}{2} -a &-b \\
d &-c &-b &\phantom{-}a
\end{array} \right ) ,
\end{equation}
with 
\begin{align}
\begin{split} \label{mat_coeff_a}
a = \frac{1}{16} \left( 3 - e^{-8\Delta^2}\left\lbrace\cos (4\theta_a) + \cos \left(4\theta_b\right)\right\rbrace \right.   
 \left. 
- e^{-16\Delta^2}\cos (4\theta_a + 4\theta_b)\right) ,
\end{split} \\
\begin{split}\label{mat_coeff_b}
b = \frac{1}{16} \left( e^{-8\Delta^2}\left\lbrace\sin (4\theta_a) - \sin \left(4\theta_b\right)\right\rbrace \right. 
 \left. - e^{-16\Delta^2}\sin (4\theta_a + 4\theta_b)\right) , \\
\end{split} \\
\begin{split} \label{mat_coeff_c}
c  = \frac{1}{16} \left( e^{-8\Delta^2}\left\lbrace - \sin (4\theta_a) + \sin \left(4\theta_b\right)\right\rbrace \right. 
 \left. 
- e^{-16\Delta^2}\sin (4\theta_a + 4\theta_b)\right) ,
\end{split} \\
\begin{split} \label{mat_coeff_d}
d = \frac{1}{16} \left(1 -  e^{-8\Delta^2}\left\lbrace\cos (4\theta_a) + \cos \left(4\theta_b\right)\right\rbrace \right. 
 \left.
+ e^{-16\Delta^2}\cos (4\theta_a + 4\theta_b)\right) .
\end{split}
\end{align}
Note that here $\Delta \rightarrow \infty$ is the \textit{completely} unsharp limit. Under this limit, the post-measurement state of equation \eqref{post} will be

\begin{equation}
\lim_{\Delta \to \infty} \rho_{ref} = \left(
\begin{array}{cccc}
\frac{3}{16} &0 &0 & \frac{1}{16}  \\
0 &\frac{5}{16} &\frac{1}{16} &0 \\
0 &\frac{1}{16} &\frac{5}{16} &0 \\
\frac{1}{16} &0 &0 &\frac{3}{16}
\end{array} \right ).
\end{equation}

\subsubsection{P-distribution}
Defining $\ket{e} \equiv \ket{\alpha_e}$ and $\ket{o} = \ket{\alpha_o}$, the P-distribution for the post-measurement state in equation (\ref{post_meas_ref_def}) will be (see appendix \ref{post_meas_cat_appendix})
\begin{equation} \label{pdist}
\mathcal{P}_{ref}(\beta, \gamma) =  \sum\limits_{i, j, k, l, \in \left\lbrace e, o\right\rbrace}\rho_{i,j,k,l}P_{ij}(\beta)P_{kl}(\gamma),
\end{equation}
where $\beta$ and $\gamma$ are the phase space variables of the two modes, and
\begin{equation}
  \rho_{i,j,k,l} = \Tr\left( \rho_{ref} \ketbra{i}{j} \otimes \ketbra{k}{l} \right) \hspace{2cm} i,j,k,l \in \{ e, o\},
 \end{equation} 
\begin{align}
\begin{split}
P_{ee}(\beta) = N_e^2   \left\lbrace 1 + e^{-2|\alpha|^2} \hat{A}(\alpha)   \right\rbrace  
 \left[ \delta^{(2)}(\alpha - \beta)  +\delta^{(2)}(\alpha + \beta)\right], 
\end{split} \\
\begin{split}
P_{eo}(\beta) = N_eN_o   \left\lbrace 1 + e^{-2|\alpha|^2} \hat{A}(\alpha)   \right\rbrace  
 \left[ \delta^{(2)}(\alpha - \beta)  - \delta^{(2)}(\alpha + \beta)\right],
\end{split} \\
\begin{split}
P_{oe}(\beta) = N_eN_o   \left\lbrace 1 - e^{-2|\alpha|^2} \hat{A}(\alpha)   \right\rbrace 
 \left[ \delta^{(2)}(\alpha - \beta)  - \delta^{(2)}(\alpha + \beta)\right],  
\end{split} \\
\begin{split}
P_{oo}(\beta) = N_o^2   \left\lbrace 1 - e^{-2|\alpha|^2} \hat{A}(\alpha)   \right\rbrace  
 \left[ \delta^{(2)}(\alpha - \beta)  + \delta^{(2)}(\alpha + \beta)\right],  
\end{split}
\end{align}
with $N_{e} = \frac{1}{\sqrt{2 + 2\exp (-2 |\alpha|^2 )}} $,  $N_o = \frac{1}{\sqrt{2 - 2\exp (-2 |\alpha|^2 )}} $, and
\begin{align}
\hat{A}(\alpha) = \sum\limits_{n = 0}^{\infty} \frac{(-1)^n}{n!}(2\alpha)^n \left( \frac{\partial}{\partial \alpha} \right)^n .
\end{align}
Note that $P_{ee}(\beta)$ corresponds to the P-distribution of the single-mode state $\alketbra{e}{e}$ and in general $P_{ij}$ corresponds to the P-distribution calculation of the term $\alketbra{i}{j}$. The P-distribution can be seen to be a highly singular function once again and hence has no classical analogue.

\subsubsection{Wigner Function}
The P-distribution by virtue of its singular nature allows us to easily identify states which are non-classical. However, due to those same reasons, it is not a well-behaved function. It is therefore illuminating to study the Wigner function of the post-measurement state as it does not have any such singularities. Here non-classicality can be understood in terms of negativity of the Wigner function, which is non-zero for any pure non-classical state which happens to be non-Gaussian. Note that, in our case, although the post-measurement state is not pure (in general), we still adopt the negativity of the Wigner function of this state to quantify its non-classicality. From the P-distribution in equation (\ref{pdist}), we can directly obtain the Wigner function \cite{cahill_gluaber}. For the state $\rho_{ref}$ in equation (\ref{post_meas_ref_def}), it is given by
\begin{equation}\label{wdist}
\mathcal{W}_{ref}(\beta, \gamma) =  \frac{1}{\pi^4}\sum\limits_{i, j, k, l \in \lbrace e, o \rbrace}\rho_{i,j,k,l } W_{ij}(\beta) W_{kl}(\gamma) ,
\end{equation}
where $\beta$ and $\gamma$ are the phase space variables and 
\begin{align}
\begin{split}
W_{ee}(\beta) = \frac{2}{\pi}N_{e}^2
\left\lbrace e^{-2 | \alpha - \beta |^2} + e^{-2 | \alpha + \beta |^2}    \right. 
\left.
+ e^{-2|\alpha|^2} \left[ e^{2(\alpha - \beta)(\alpha^* + \beta^* )}  + e^{2(\alpha + \beta)(\alpha^* - \beta^* )} \right] \right\rbrace ,
\end{split} \\
\begin{split}
W_{eo}(\beta) = \frac{2}{\pi}N_{e}N_o
\left\lbrace e^{-2 | \alpha - \beta |^2} - e^{-2 | \alpha + \beta |^2}    \right. 
\left.
+ e^{-2|\alpha|^2} \left[ e^{2(\alpha - \beta)(\alpha^* + \beta^* )}  - e^{2(\alpha + \beta)(\alpha^* - \beta^* )} \right] \right\rbrace ,
\end{split} \\
\begin{split}
W_{oe}(\beta) = \frac{2}{\pi}N_{e}N_o
\left\lbrace e^{-2 | \alpha - \beta |^2} - e^{-2 | \alpha + \beta |^2}    \right. 
\left.
- e^{-2|\alpha|^2} \left[ e^{2(\alpha - \beta)(\alpha^* + \beta^* )}  - e^{2(\alpha + \beta)(\alpha^* - \beta^* )} \right] \right\rbrace ,
\end{split} \\
\begin{split}
W_{oo}(\beta) = \frac{2}{\pi} N_o^2
\left\lbrace e^{-2 | \alpha - \beta |^2} + e^{-2 | \alpha + \beta |^2}    \right. 
\left.
- e^{-2|\alpha|^2} \left[ e^{2(\alpha - \beta)(\alpha^* + \beta^* )}  + e^{2(\alpha + \beta)(\alpha^* - \beta^* )} \right] \right\rbrace .
\end{split} 
\end{align}
Note that $\mathcal{W}_{ref}(\beta, \gamma )$ is not a Guassian function. Similar to the P-distribution, $W_{ij}$, $i$, $j \in \lbrace{e, o \rbrace}$, represents the Wigner function of the operator $\alketbra{i}{j}$. The effect of reference coarse-graining is reflected in the paramters $a$, $b$, $c$ and $d$ (given respectively by equations, \eqref{mat_coeff_a}, \eqref{mat_coeff_b}, \eqref{mat_coeff_c} and \eqref{mat_coeff_d}) which depend on the coarse-graining parameter $\Delta$.

\subsubsection{Negativity as a measure of non-classicality}
For any two-mode state $\rho$, it is known that, the negativity of the phase space volume of the Wigner function, given by
\begin{equation}
\mathcal{N}_{\rho} = \frac{1}{2}\int \left( |\mathcal{W} ( \beta, \gamma ) | - \mathcal{W}(\beta, \gamma ) \right) \Diff2 \beta \Diff2 \gamma, 
\end{equation}
can be used as an indicator of non-classicality of the state $\rho$ \cite{kenfack2004negativity}. While there are still open questions about which aspects of non-classicality is captured by this measure \cite{ferraro2012nonclassicality}, there is evidence that it captures non-Guassianity, entanglement, and correlations beyond entanglement such as discord \cite{marek2009nonclassicality} \cite{taghiabadi2016}.
\begin{figure}[!]
\centering
\includegraphics[width = 0.48\textwidth]{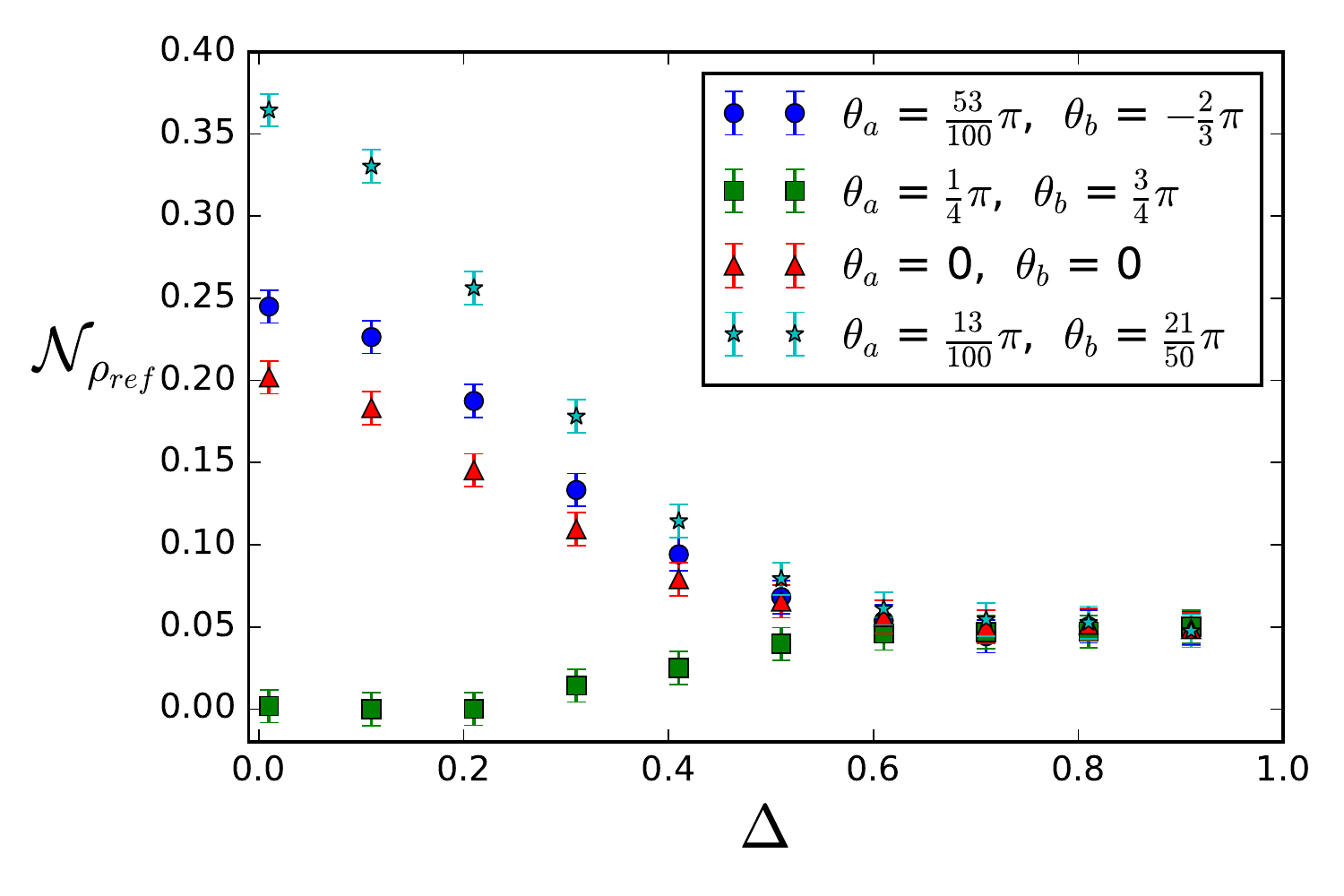} 
\caption{Wigner function negativity of the post-measurement state of $\ket{\psi_{\alpha}}$ for different rotations of the reference coarse-grained measurement operator $O_{\Delta}(\theta_{a}, \theta_{b})$, with $\alpha = 2$. (color online). }
\label{fig_ref}
\end{figure}

\begin{figure}
\centering
\begin{minipage}{0.48\textwidth}
\includegraphics[width = \linewidth]{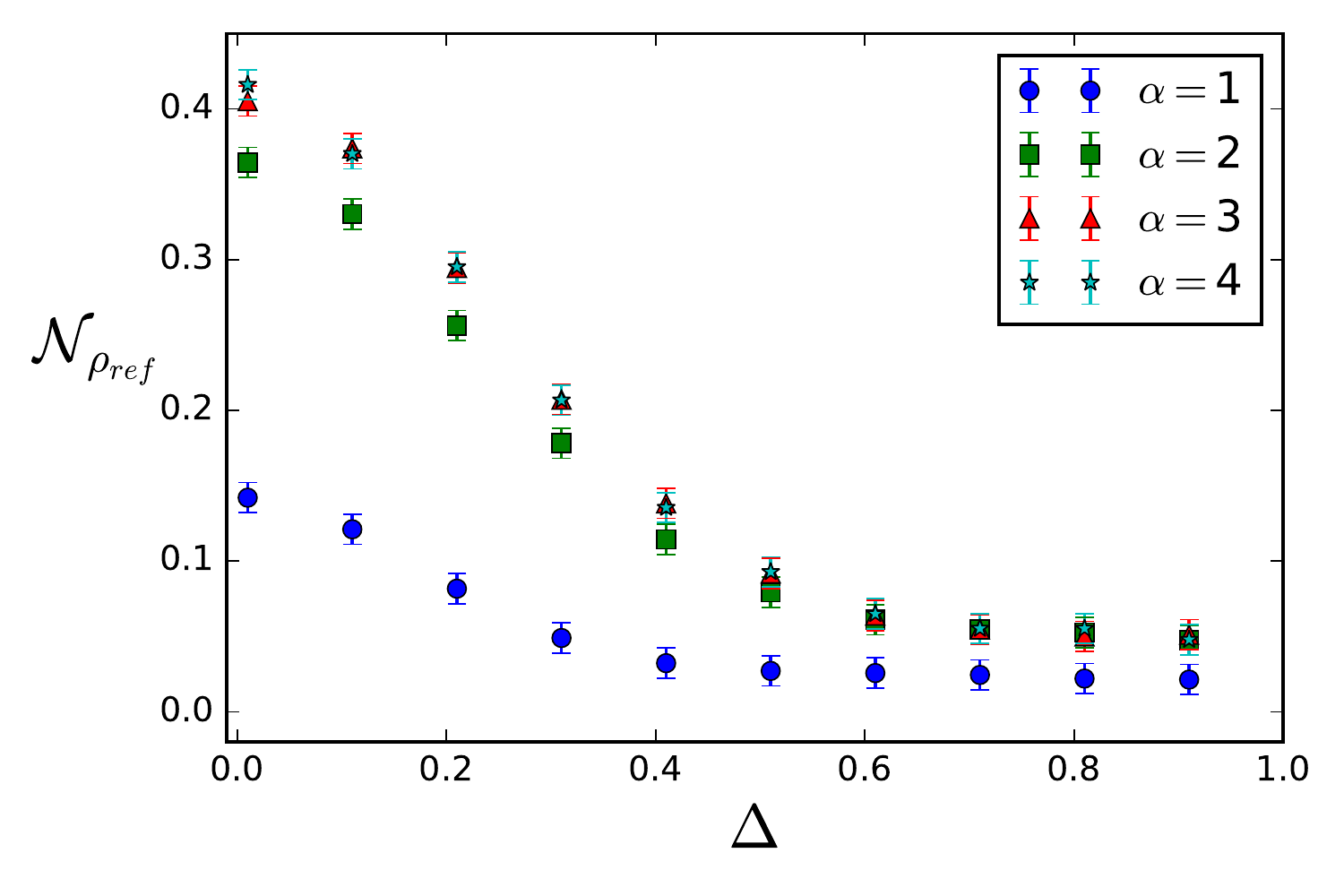}
\caption{Wigner function negativity of the post-measurement state of $\ket{\psi_{\alpha}}$ after the reference coarse-grained measurement $O_{\Delta}(\frac{13\pi}{100}, \frac{21\pi}{50} )$ (color online). }
\label{aplot_041_132_plot}
\end{minipage}
\hfill
\begin{minipage}{0.48\textwidth}
\includegraphics[width = \linewidth]{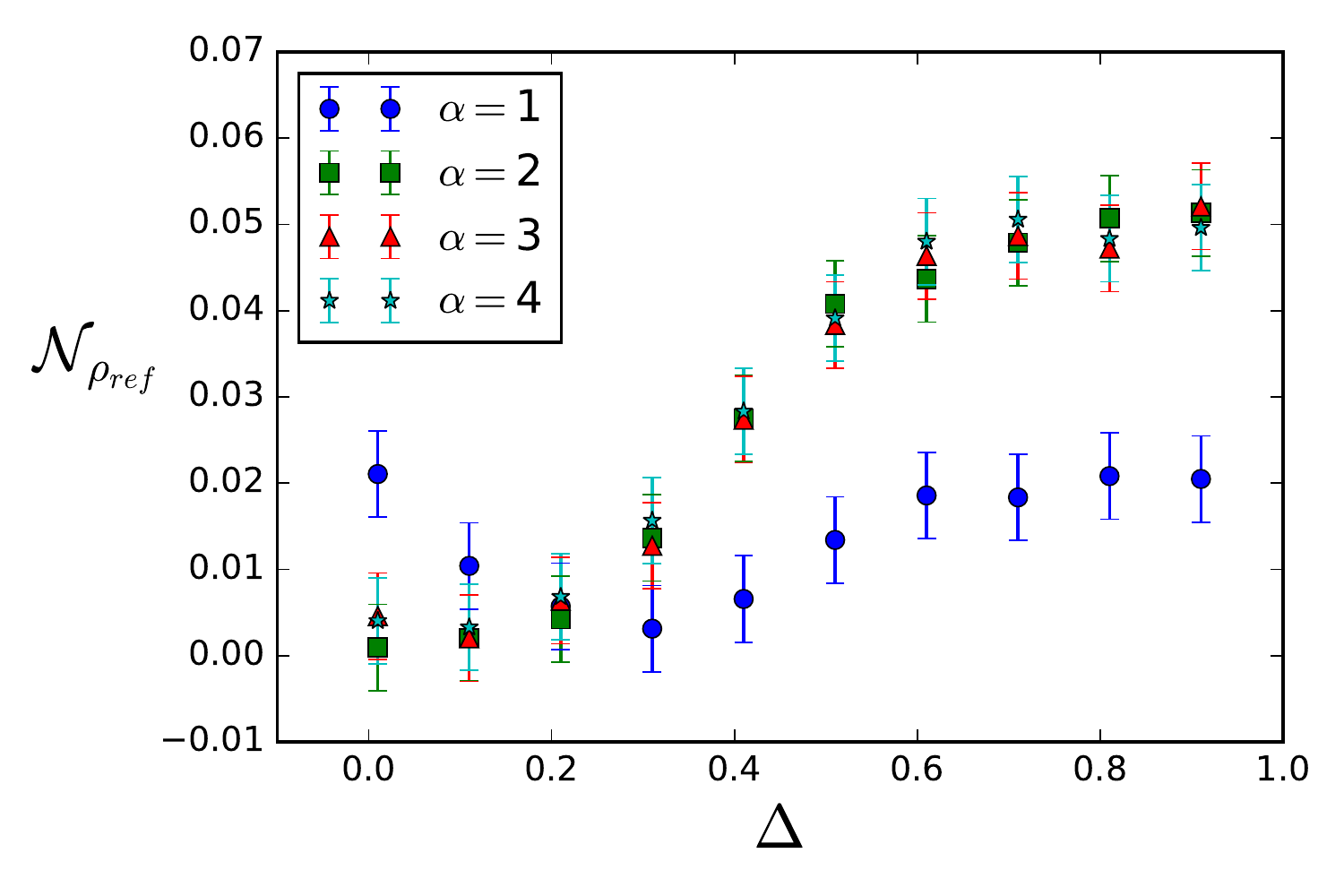}
\caption{Wigner function negativity of the post-measurement state of $\ket{\psi_{\alpha}}$ after the reference coarse-grained measurement $O_{\Delta}(\frac{\pi}{4}, \frac{3\pi}{4})$ (color online). }
\label{aplot_079_236_plot}
\end{minipage}
\end{figure}

The negativity of $\mathcal{W}_{ref}( \beta, \gamma )$ as a function of the reference coarse-graining parameter $\Delta$, is plotted for different values of the rotation angles $\theta_{a}$ and $\theta_{b}$ \footnote{For all $\theta_a$ and $\theta_{b}$ chosen, it is possible to find $\theta_{c}$, $\theta_{d}$ such that for a sharp measurement, the CHSH-Bell quantity in equation \eqref{bell_quanity_def} is greater than 2. } of the measurement operator $O_{\Delta} (\theta_{a}, \theta_{b})$, given in equation (\ref{ref_twomode_meas_def}), in FIG. \ref{fig_ref}. Depending on the choice of $\theta_{a}$ and $\theta_{b}$, the post-measurement state after sharp measurement ($\Delta = 0$) can have different values of negativity. But coarse-graining does not always decrease the negativity, in fact, with the measurement operator $O_{\Delta}(\theta_{a} = \frac{\pi}{4} ,\theta_{b} = \frac{3\pi}{4}) $, the negativity of the post-measurement state Wigner fuction $\mathcal{W}_{ref}$ can be seen to be increased to a non-zero value with increasing coarse-graining parameter $\Delta$. The dependence of $\mathcal{N}_{ref}$ on $\alpha$ can be seen in FIGs. \ref{aplot_041_132_plot} and \ref{aplot_079_236_plot}; in general, the negativity increases with $\alpha$ but there are exceptions such as the $\alpha = 1$ case in FIG. \ref{aplot_079_236_plot}.
\subsection{Resolution coarse-graining}

\begin{figure}
\begin{minipage}{0.48\textwidth}
\includegraphics[width = \linewidth]{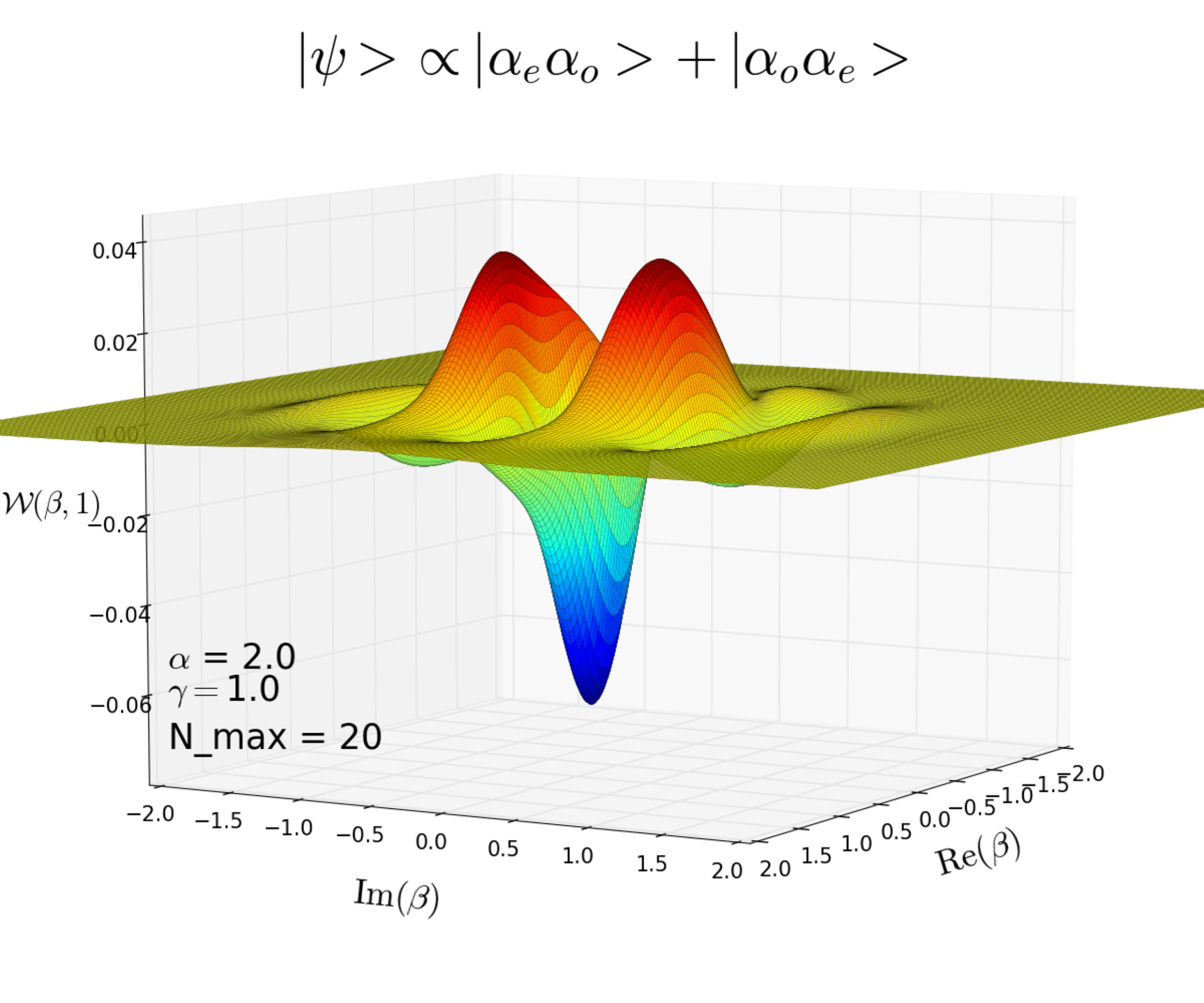}
\caption{Wigner function $\mathcal{W}_{\psi_{\alpha}}(\beta, \gamma)$ of $\ket{\psi_{\alpha}}$ with $\gamma$ constant, N\_max is the maximum number of photons in each mode (color online).}
\label{plt_initial}
\end{minipage}
\hfill
\begin{minipage}{0.48\textwidth}
\includegraphics[width = \linewidth]{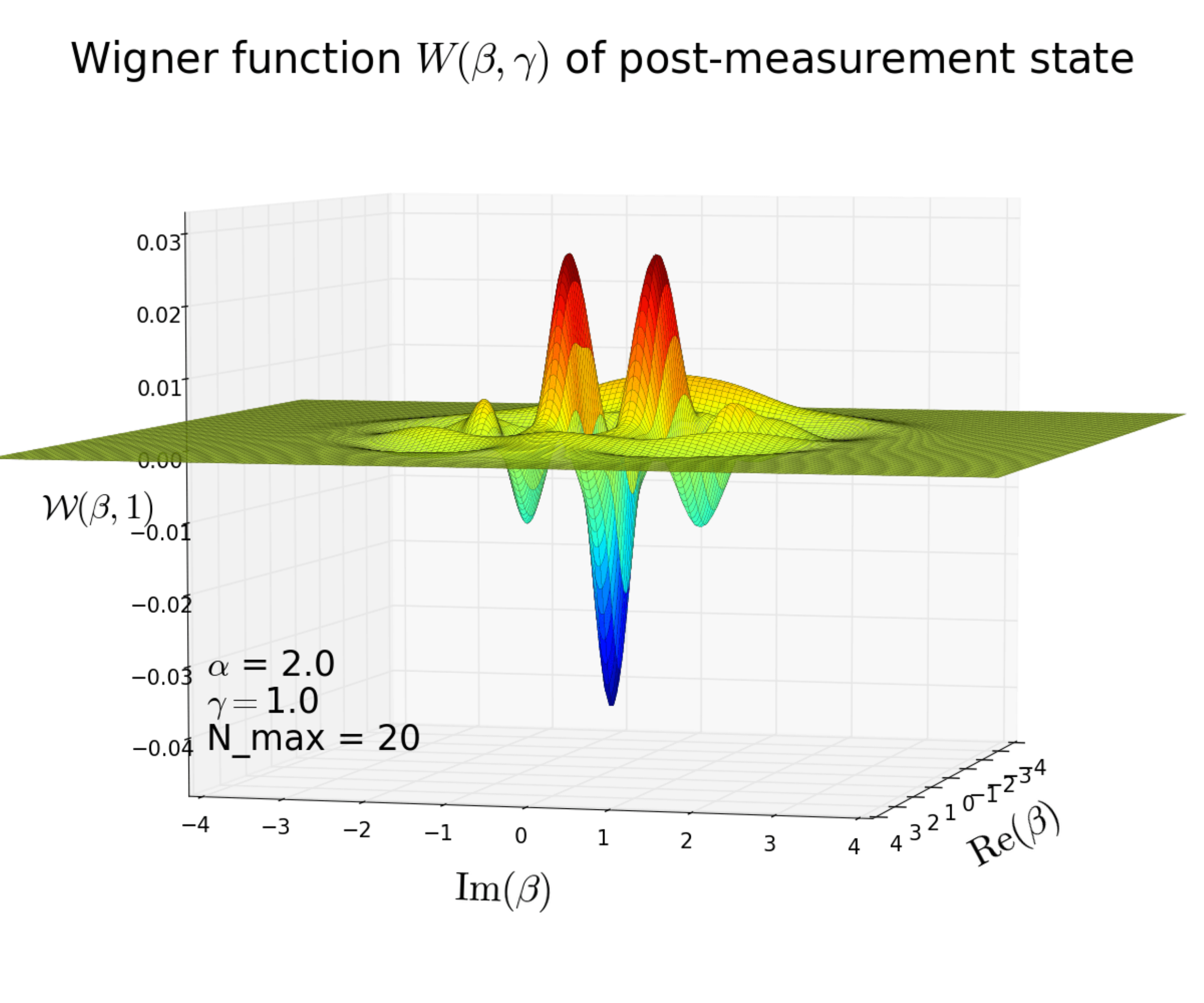}
\caption{Wigner function $\mathcal{W}(\beta, \gamma)$ of the post-measurement state of $\ket{\psi_{\alpha}}$ after sharp measurement with $O^{4}(\frac{\pi}{4}) \otimes O^{4}(\frac{3\pi}{4})$, N\_max is the maximum number of photons in each mode (color online).}
\label{plt_sharp}
\end{minipage}
\end{figure}

Let us now turn to resolution coarse-graining with the measurement operator being of the form
\begin{equation}
O_{\delta}(\theta_a, \theta_b) = \sum\limits_{k, m = -\infty}^{+\infty}P_{\delta}(k)P_{\delta}(m)O^{k}(\theta_a) \otimes O^{m}(\theta_b),
\end{equation}
where $P_{\delta}(k)$s are again discrete Guassians as before. For the initial state $\ket{\psi_{\alpha}} = \frac{1}{\sqrt{2}}\left( \ket{\alpha_e \alpha_o} + \ket{\alpha_o \alpha_e} \right)$, the post-measurement state for non-selective measurement is given by
\begin{equation}
\begin{split}
\rho_{res} = \sum\limits_{k, m = -\infty}^{+\infty}P_{\delta}(k)P_{\delta}(m) \left\lbrace \phantom{O^{k}_{+}(\theta_a)} \right. \\ \left.
O^{k}_{+}(\theta_a)O^{m}_{+}(\theta_b) |\psi_{\alpha}\rangle\langle \psi_{\alpha} | O^{k}_{+}(\theta_a)O^{m}_{+}(\theta_b) \right. \\ \left.
+ O^{k}_{+}(\theta_a)O^{m}_{-}(\theta_b) |\psi_{\alpha}\rangle\langle \psi_{\alpha} | O^{k}_{+}(\theta_a)O^{m}_{-}(\theta_b) \right. \\ \left.
+ O^{k}_{-}(\theta_a)O^{m}_{+}(\theta_b) |\psi_{\alpha}\rangle\langle \psi_{\alpha} | O^{k}_{-}(\theta_a)O^{m}_{+}(\theta_b) \right. \\ \left.
+ O^{k}_{-}(\theta_a)O^{m}_{-}(\theta_b) |\psi_{\alpha}\rangle\langle \psi_{\alpha} | O^{k}_{-}(\theta_a)O^{m}_{-}(\theta_b)  \right\rbrace.
\end{split}
\end{equation}
Since we are interested in the effect of resolution coarse-graining, we fix our basis by choosing $(\theta_a, \theta_b) = (\frac{\pi}{4} , \frac{3\pi}{4})$. The Wigner function $\mathcal{W}_{\psi_{\alpha}}(\beta, \gamma)$ of the initial state $\ket{\psi_{\alpha}}$ is plotted in FIG. \ref{plt_initial}. $\gamma$ is kept constant while $\beta$  is varied. The Fock space of each mode was truncated at a maximum number state of $|20\rangle$.
The post-measurement state for a sharp measurement, using the operator $O^{k = 4}(\theta_a = \frac{\pi}{4})\otimes O^{k = 4}(\theta_b = \frac{3\pi}{4})$ is plotted in FIG. \ref{plt_sharp}. We have chosen $\gamma$ and $k$ such that this state can be clearly seen to have negativity.
\begin{figure}[!]
\centering
\begin{minipage}{0.48\textwidth}
\includegraphics[width = \linewidth]{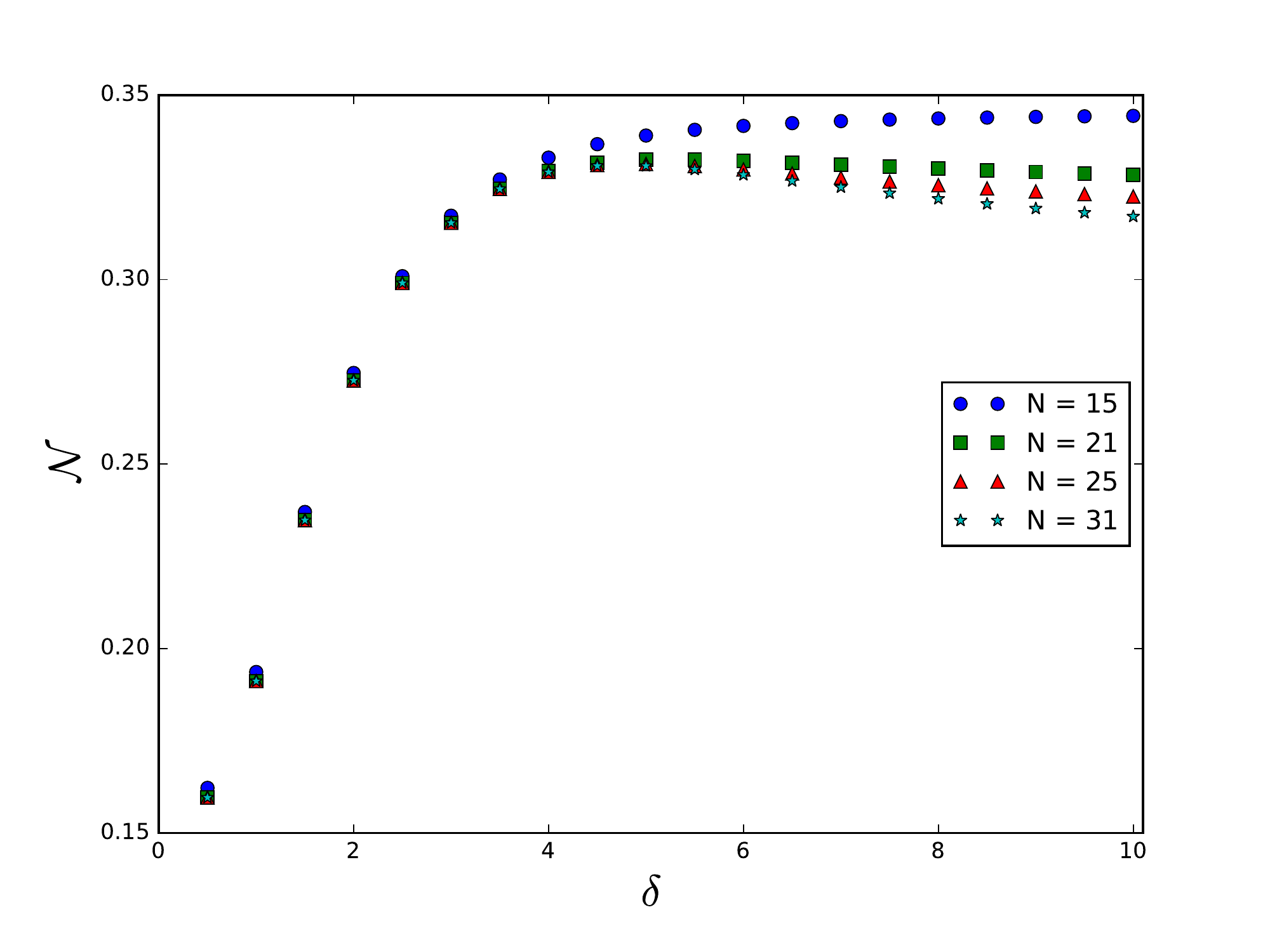}
\caption{Wigner function negativity of the post-measurement state of $\ket{\psi_{\alpha}}$ vs. the resolution coarse-graining parameter after measurement with $O_{\delta}(\frac{\pi}{4}, \frac{3\pi}{4})$ for different truncations N, with $\alpha = 2$ (color online).}
\label{negvsdelta}
\end{minipage}
\hfill
\begin{minipage}{0.48\textwidth}
\includegraphics[width = \linewidth]{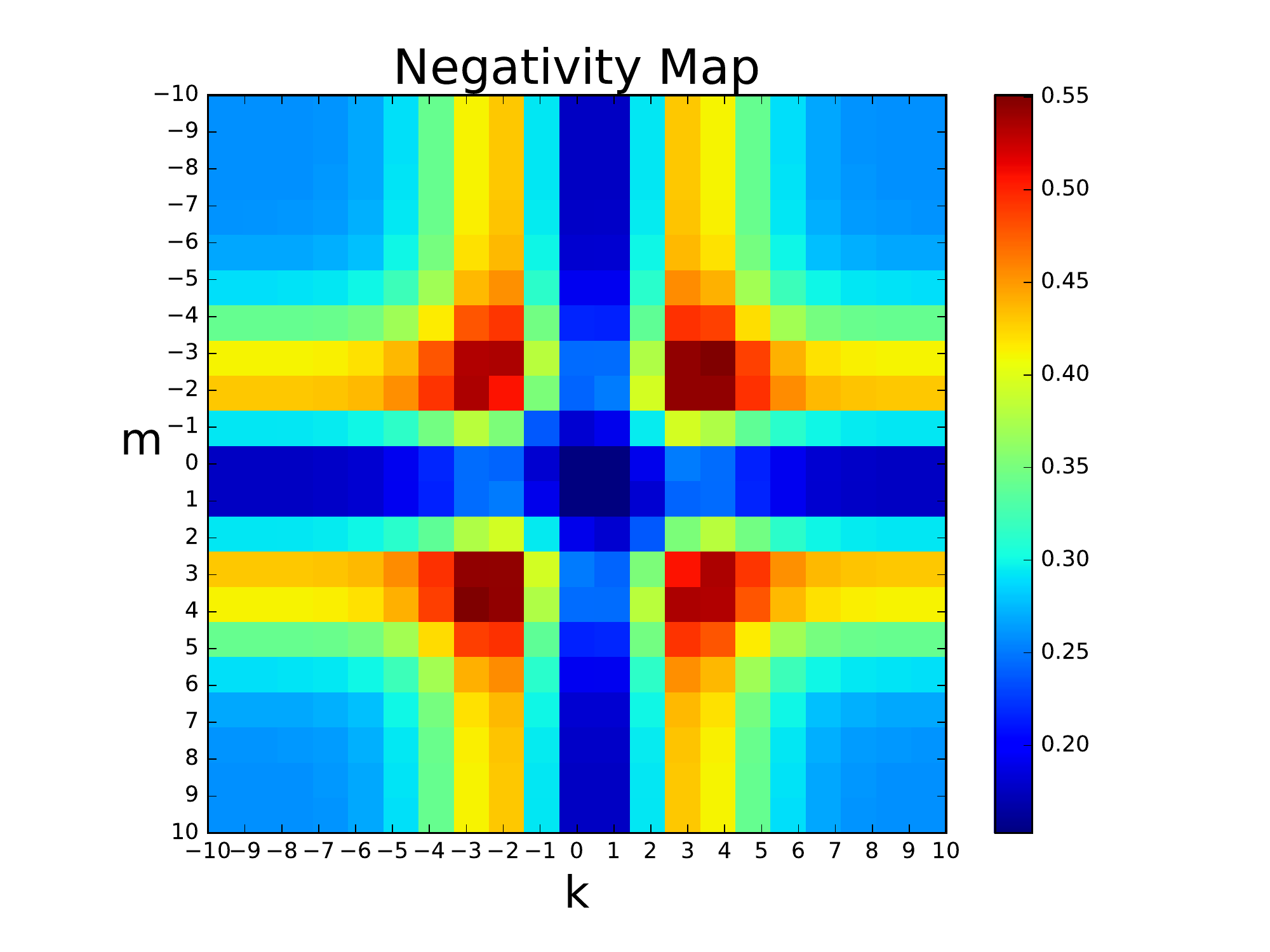}
\caption{Wigner function negativity of the post-measurement state of $\ket{\psi_{\alpha}}$ after sharp measurement with $O^{k}(\frac{\pi}{4}) \otimes O^{m}(\frac{3\pi}{4})$ for different values of $k$ and $m$ (colour online).}
\label{neg_map}
\end{minipage}
\end{figure}
Negativity vs. the resolution coarse-graining parameter $\delta$ is plotted for different values of Fock space truncation in FIG. \ref{negvsdelta}. Contrary to the Bell-type inequality results, here non-classicality is seen to be increasing with $\delta$. This result can be understood by looking at the negativity of sharp measurements $O^{k}(\theta_a) \otimes O^{m}(\theta_b)$ with respect to $k$ and $m$. Let $N_{km}(\theta_a , \theta_b)$ be the negativity of the Wigner function of the post-measurement state with sharp measurement $O^{k}(\theta_a)\otimes O^{k}(\theta_b)$. The negativity of the Wigner function for the post-measurement state $\rho_{res}$ can be written as
\begin{equation}\label{neg_res_eq_def}
\mathcal{N}_{\rho_{res}} = \sum\limits_{k, m = -\infty}^{+\infty} P_{\delta}(k)P_{\delta}(m)N_{km}.
\end{equation}
The value of $N_{km}(\theta_a = \frac{\pi}{4}, \theta_b = \frac{3\pi}{4})$ is plotted against $k$ and $m$ in FIG. \ref{neg_map}. The $\delta \rightarrow 0$ case corresponds to the measurement  $O^{0}(\theta_a = \frac{\pi}{4}) \otimes O^{0}(\theta_b = \frac{3\pi}{4})$. The negativity of the Wigner function of the post-measurement state, corresponding to this measurement, is smaller than that of the surrounding post-measurement states with measurements corresponding to $k \neq 0$, $m \neq 0$. When $\delta$ is increased, contributions to the sum in equation (\ref{neg_res_eq_def}) from terms of larger negativity with higher values of $k$ and $m$  increase and this leads to an increase in the overall post-measurement state negativity $\mathcal{N}_{\rho_{res}}$.

\section{Measurement of NOON states} \label{ent_fock_sec}
\begin{figure}
\centering
\begin{minipage}{0.48\textwidth}
\includegraphics[width = \linewidth]{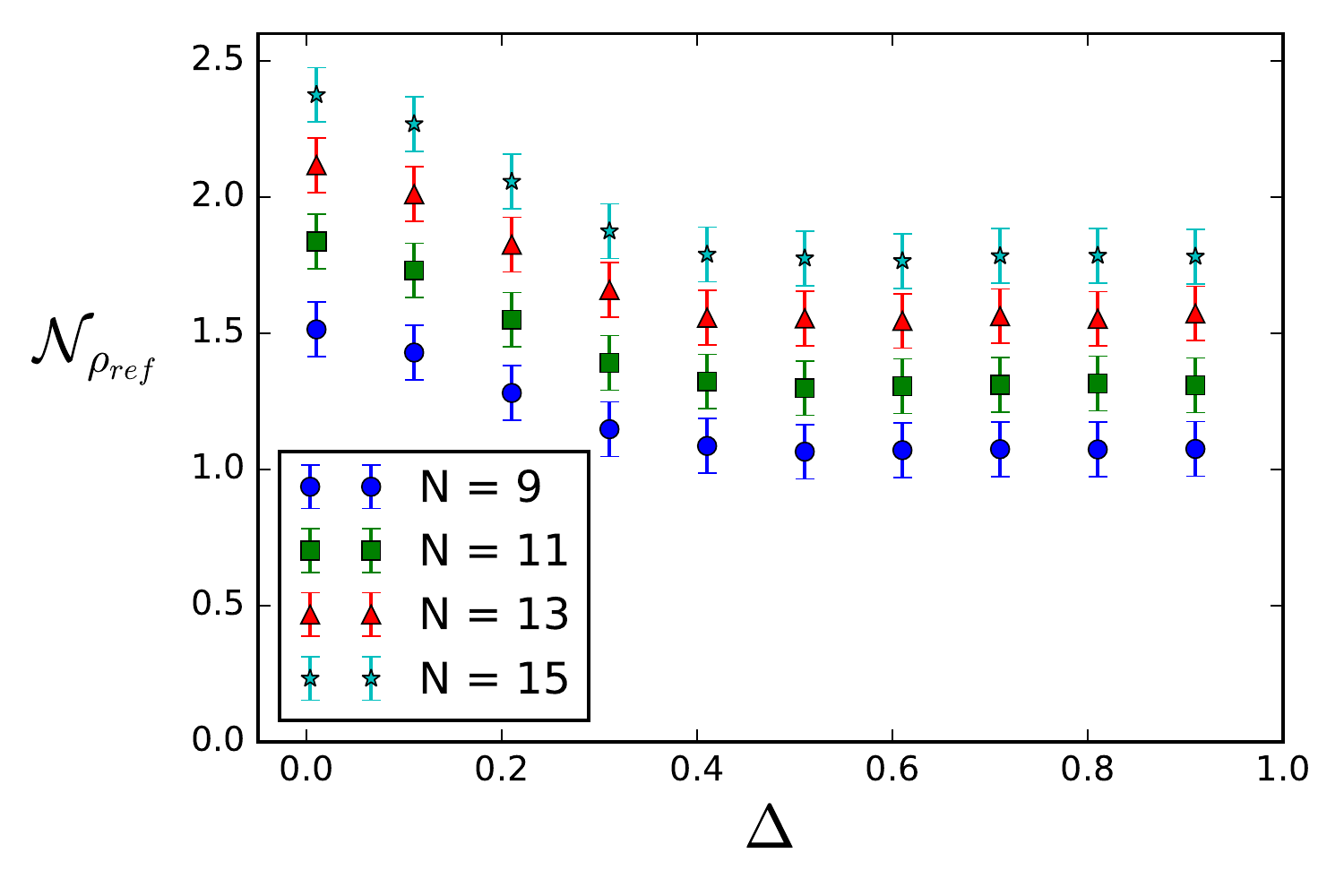}
\caption{Wigner function negativity of the post-measurement state of $\ket{\psi_{N}}$ after the reference coarse-grained measurement with $O_{\Delta}(\frac{13\pi}{100}, \frac{21\pi }{50})$ (color online).}
\label{0n_041_132_plot}
\end{minipage}
\hfill
\begin{minipage}{0.48\textwidth}
\includegraphics[width = \linewidth]{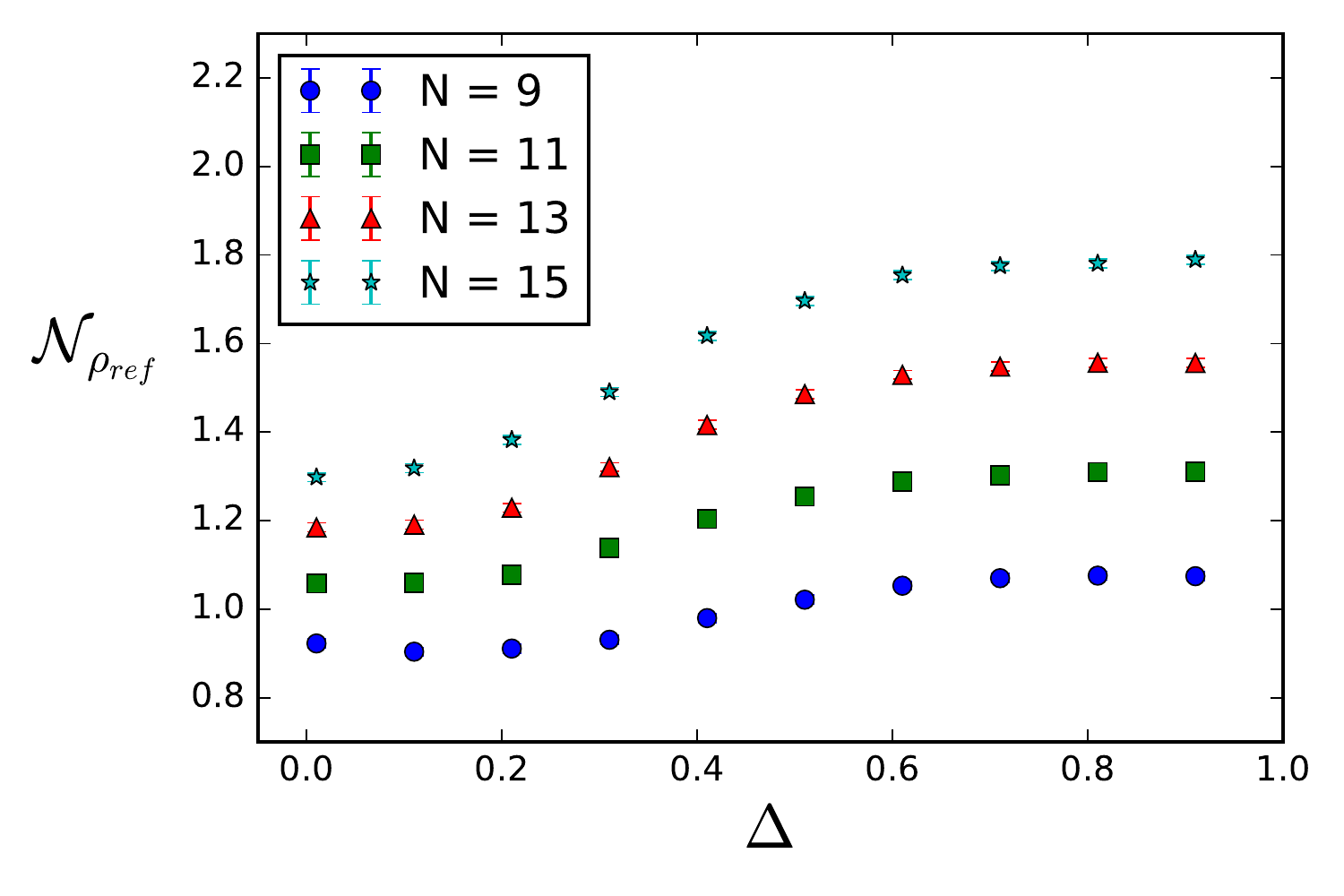}
\caption{Wigner function negativity of the post-measurement state of $\ket{\psi_{N}}$ after the reference coarse-grained measurement $O_{\Delta}(0, 0)$ (color online).}
\label{0n_00_00_plot}
\end{minipage}
\end{figure}

Note that the expression for the Bell quantity under reference coarse-graining given by equation  \eqref{bell_Delta_eqn} remains unchanged for a general class of initial states obtained by the substitution $\ket{\alpha_e} \rightarrow \ket{e}$ and $\ket{\alpha_o} \rightarrow \ket{o}$, where
\begin{equation}\label{odd_even_form_def}
\ket{e} \equiv \sum\limits_{n = 0}^{+\infty} C_{2n} \ket{2n}, \hspace{3cm} \ket{o} \equiv \sum\limits_{n = 0}^{+\infty} C_{2n + 1} \ket{2n + 1},
\end{equation}
i.e., $\ket{e}$ has support only on even Fock states and $\ket{o}$ has support on only odd Fock states, as long as they are related by the unitary operator as,
\begin{equation}
\begin{split}
&U(\theta) \ket{e} = \cos \theta \ket{e} + \sin\theta \ket{o}, \\
&U(\theta) \ket{o} = \sin \theta \ket{e} - \cos\theta \ket{o}. 
\end{split}
\end{equation}
To see a trade-off between initial state non-classicality and post-measurement state negativity, it is illuminating to choose the even and odd states to be Fock states such that $\ket{e} = \ket{0}$ and $\ket{o} =  \ket{2n + 1}$, n $\in$ $\lbrace 0, 1, 2,...\rbrace$. Note that the initial state will now be the NOON state
\begin{equation}
\ket{\psi_{N}} = \frac{1}{\sqrt{2}}\left( \ket{0}\ket{N} + \ket{N}\ket{0}  \right), \hspace{3cm} N \in \lbrace 1, 3, 5,... \rbrace.
\end{equation}
This choice leaves the expression for $B_{\Delta}$ in equation \eqref{bell_Delta_eqn} unchanged by satisfying equation \eqref{odd_even_form_def}. The Wigner function of post-measurement state $\rho_{ref}$ under the reference coarse-grained measurement, defined by the operator in equation \eqref{ref_twomode_meas_def}, will be given again by the expression in equation \eqref{wdist}, with (see appendix \ref{ent_fock_appendix})
\begin{align}
\begin{split}
W_{ee}(\beta) = \frac{2}{\pi} e^{- 2 |\beta|^2} ,
\end{split} \\
\begin{split}
W_{eo}(\beta) = \frac{2}{\pi} \frac{(2 \beta)^{N}}{\sqrt{N !}}e^{- 2 |\beta |^2},
\end{split} \\
\begin{split}
W_{oe}(\beta) = \frac{2}{\pi} \frac{(2 \beta^*)^{N}}{\sqrt{N!}}e^{- 2 |\beta|^2},
\end{split} \\
\begin{split}
W_{oo}(\beta) = -\frac{2}{\pi} L_{N}(4|\alpha|^2),
\end{split} 
\end{align}
where $L_{n}(x)$ is the $n^{th}$ Laguerre polynomial. For typical values of the unitary rotation angles $\theta_1$ and $\theta_2$ it is found that negativity of the Wigner function of $\rho_{ref}$ decreases with increase in the coarse-graining parameter $\Delta$ as shown in FIG. \ref{0n_041_132_plot} for $(\theta_1, \theta_2 ) = (\frac{13}{100}\pi, \frac{21}{50}\pi )$. However this is not strictly true. For the choice $(\theta_1, \theta_2 ) = (0, 0 )$, it can be seen that negativity increases with $\Delta$ as seen in FIG. \ref{0n_00_00_plot}. But in all cases, the negativity of the Wigner function increases with photon number $N$.

\begin{figure}
\centering
\begin{minipage}{0.48\textwidth}
\includegraphics[width = \linewidth]{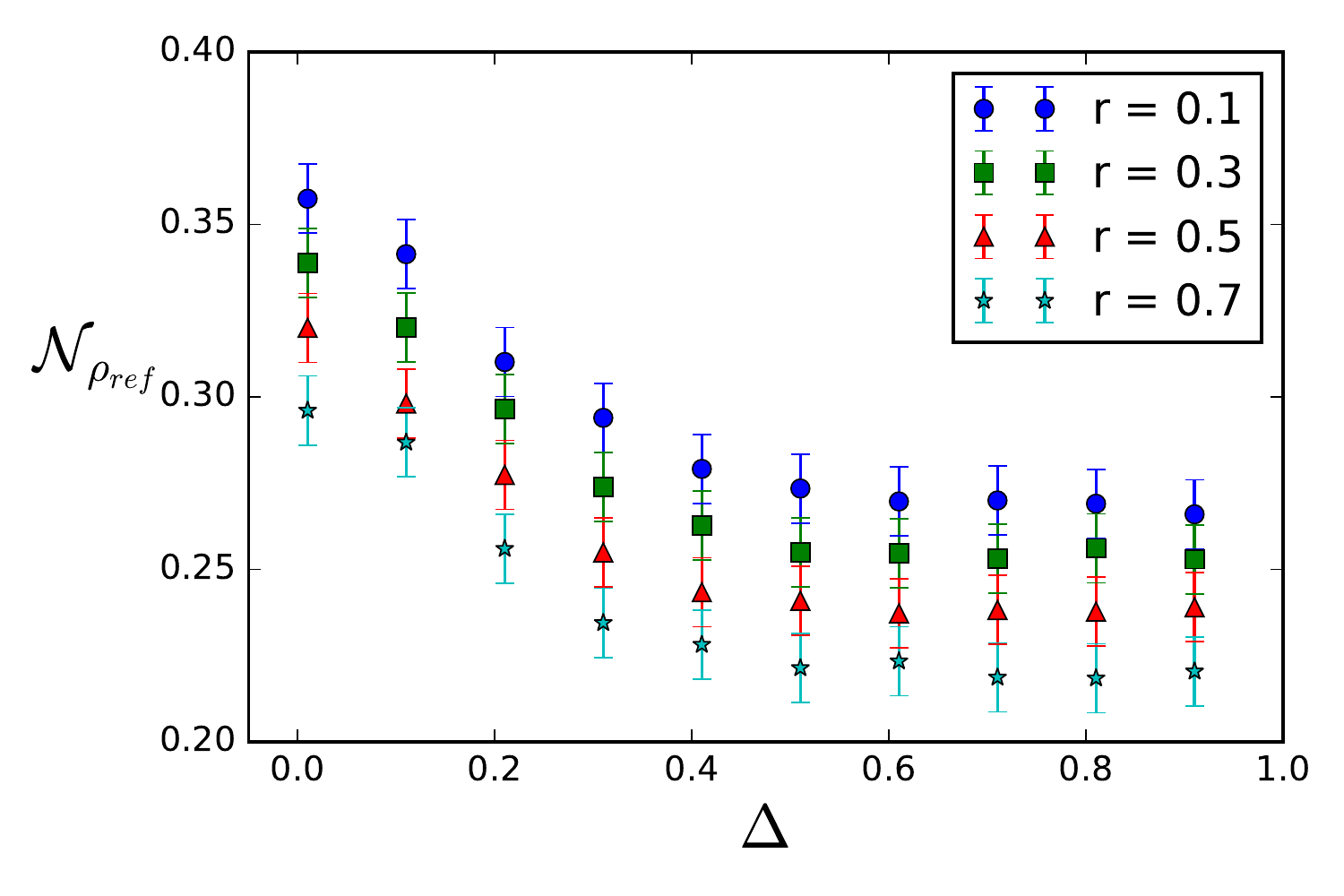}
\caption{Wigner function negativity of the post-measurement state of $\ket{\psi_{r}}$ after the reference coarse-grained measurement $O_{\Delta}(\frac{53\pi}{100}, \frac{2\pi}{3} )$ (color online).}
\label{r_167_-207_plot}
\end{minipage}
\hfill
\begin{minipage}{0.48\textwidth}
\includegraphics[width = \linewidth]{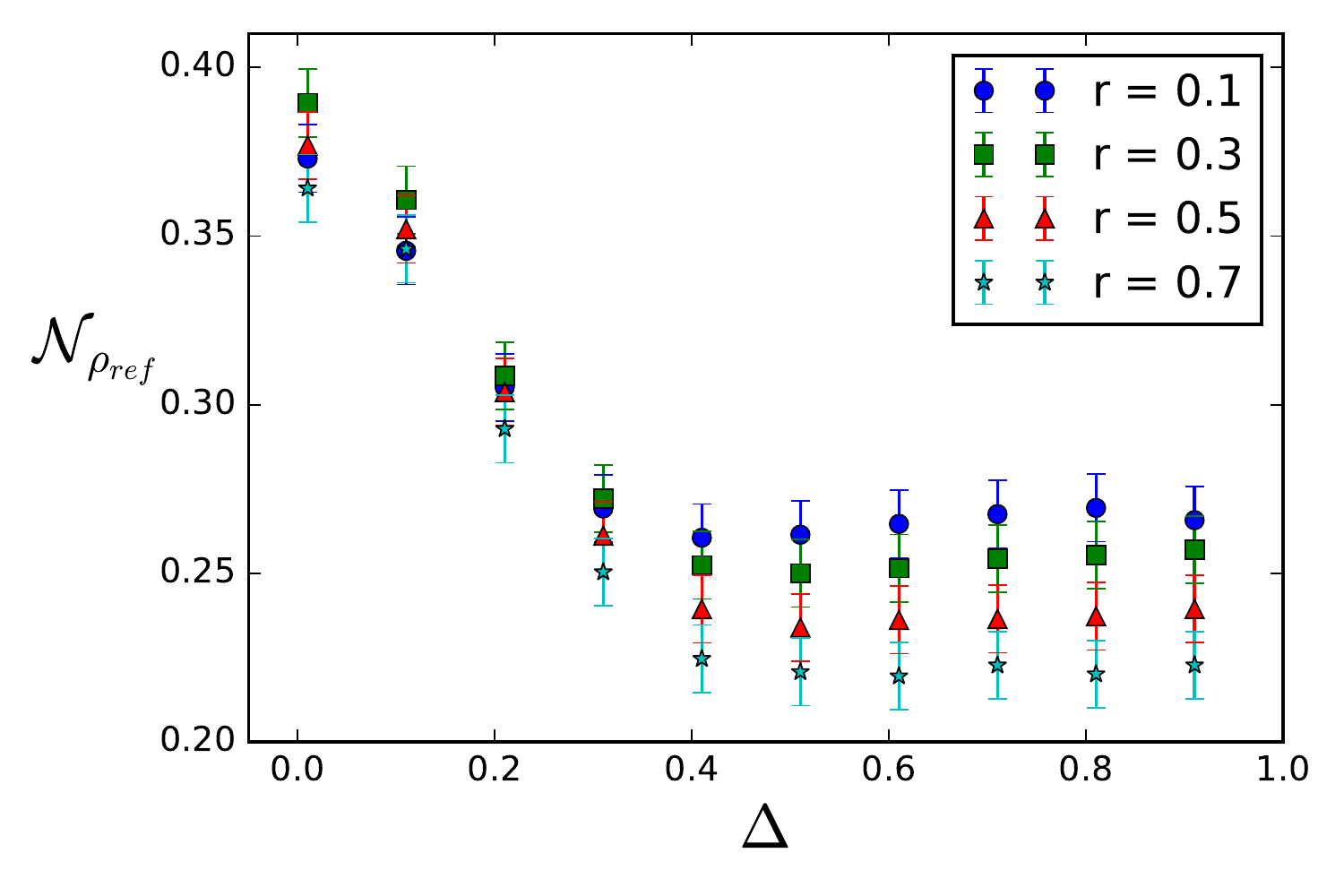}
\caption{Wigner function negativity of the post-measurement state of $\ket{\psi_{r}}$ after the reference coarse-grained measurement $O_{\Delta}(\frac{\pi}{4}, \frac{3\pi}{4})$ (color online).}
\label{r_079_236_plot}
\end{minipage}
\end{figure}

\section{Measurement of entangled photon-added squeezed vacuum}\label{squeezed_sec}
In the previous section we could see the trade-off between increased non-classicality of the initial state in the sense of higher photon number and the reference coarse-graining parameter $\Delta$. Another notion of non-classicality that is worth studying in this context is squeezing. Consider the initial state 
\begin{equation}
\label{suqueezed_initial_def}
\ket{\psi_{r}} = \frac{1}{\sqrt{2}} \left( \ket{\Psi_{+}^{r}}\ket{\Psi_{-}^{r}} + \ket{\Psi_{-}^{r}} \ket{\Psi_{+}^{r}} \right),
\end{equation}
where $\ket{\Psi_{+}^{r}}$ and $\ket{\Psi_{-}^{r}}$ are two-photon-added and one-photon-added squeezed vacuum states, respectively. They are defined as
\begin{align}
\ket{\Psi_{+}^{r}} &= \frac{1}{\cosh^2 r\sqrt{2 + \tanh^2 r}}(\hat{a}^{\dagger})^2 S(r)\ket{0}, \\
\ket{\Psi_{-}^{r}} &= \frac{1}{\cosh r }\hat{a}^{\dagger} S(r)\ket{0},
\end{align}
and $S(r) = \exp \left( \frac{r}{2}(\hat{a}^2 - (\hat{a}^{\dagger})^2 ) \right) $ is the squeezing operator with real squeezing parameter $r$. The squeezed vacuum state $S(r)\ket{0}$ has support only on even Fock states \cite{gerry_knight}; this ensures that $\ket{\Psi_{+}^{r}}$ and $\ket{\Psi_{-}^{r}}$  are even and odd states respectively, satisfying equation \eqref{odd_even_form_def} thus ensuring $B_{\Delta}$ is given by equation \eqref{bell_Delta_eqn}. The Wigner function of the post-measurement state can be computed as before (see appendix \ref{squeezed_appendix}) and the dependence of the negativity of the Wigner function on reference coarse-graining parameter $\Delta$ is given in FIGs. \ref{r_167_-207_plot} and \ref{r_079_236_plot}. Surprisingly, the relation between non-classicality of the initial state and Wigner function negativity of the post-measurement state is reversed. For higher values of the squeezing parameter, the post-measurement state negativity decreases. Note that this is again not strictly true, there are cases where for some values of $\Delta$ the ordering is not maintained as can be seen in FIG. \ref{r_079_236_plot}.

\section{ Summary and  Conclusions } \label{conclusions}

Even though the violation of any local realistic inequality does indicate some `non-classical' behaviour of the bipartite state at hand, the non-violation does not guarantee any `classicality' unless we can provide a local realistic model of the state. As the notion of non-classicality is a well-established feature in quantum optics, we looked at the P-distribution as well as the negativity of the Wigner function of the post-measurement state in the case of non-selective measurement involving both reference as well as resolution coarse-graining, by starting from some suitably chosen two-mode entangled states. Contrary to general indication of quantum-to-classical transition via non-violation of the Bell-CHSH inequality, we found the presence of non-classicality in the post-measurement states irrespective of whether we choose reference or resolution coarse-graining.

We found that the non-classicality behaviour of the post-measurement state, in the sense of negativity of the Wigner function, under reference coarse-graining, depends on the choice of measurement operator. For suitable choices of  $\theta_{a}$ and $\theta_{b}$ of the reference coarse-grained measurement operator $O_{\Delta}(\theta_{a}, \theta_{b})$, the negativity can decrease or increase with the reference coarse-graining parameter $\Delta$. For odd-even coherent states, the negativity $N_{ref}$ in general increases with $\alpha$, the complex parameter of the coherent states, with some exceptions such as the $\alpha = 1$ case in FIG. \ref{aplot_079_236_plot}. This dependence on $\theta_{a}$ and $\theta_{b}$ of the behaviour of negativity versus coarse-graining parameter $\Delta$ is seen again with NOON states as well as entangled photon-added squeezed vacuum states. For NOON states, $N_{ref}$ increases with photon number as might be expected because larger photon number states are regarded as more non-classical. However, for the entangled photon-added squeezed vacuum states, the post-measurement Wigner function negativity is seen to decrease with increasing value of the squeezing parameter $r$, suggesting that the negativity of the Wigner function is limited in capturing this aspect of non-classicality. As seen from FIG. \ref{r_079_236_plot}, lower value of r does not lead to higher negativity for all values of $\Delta$ but does so asymptotically in the limit of large $\Delta$. To compare the three cases presented in this paper, it will be pertinent to study the behavior of non-classicality of the post-measurement state with respect to average photon number in the initial state $\braket{\psi_{i} | N_1 \otimes \mathbb{I}+ \mathbb{I} \otimes N_2 | \psi_{i}}$, where, $N_{1}$ and $N_{2}$ are single mode photon number operators, for a fixed value of the coarse-graining parameter. In the future, we would like to consider also the study of resolution coarse-graining for the initial states considered in sections \ref{ent_fock_sec} as well as \ref{squeezed_sec}. We would like to determine the effect of rotations more general than what has been considered here for reference coarse-graining to study quantum-to-classical transition. The present study can, in principle, be extended to multipartite systems. We are also interested in relating our analysis with other approaches to quantum-to-classical transitions such as the one put forth in \cite{gisin_2014}. We hope that our study will help better understand the notion of quantum-to-classical transition.

\section{Acknowledgments}   

TB would like to thank Shankar G. Menon, Suman Mondal and Karunnya Dhevi for initial discussions on the problem of quantum-to-classical transition when all of them pursued their summer internship at IMSc, Chennai, in the summer of 2015. TB acknowledges the hospitality of IMSc for pursuing the aforementioned summer programme, during which part of the work was done. SG acknowledges useful discussions with Samir Kunkri about the issue of quantum-to-classical transition.

\nocite{*}
\bibliographystyle{apsrev}
\bibliography{refs}        

\begin{thebibliography}{20}
\expandafter\ifx\csname natexlab\endcsname\relax\def\natexlab#1{#1}\fi
\expandafter\ifx\csname bibnamefont\endcsname\relax
  \def\bibnamefont#1{#1}\fi
\expandafter\ifx\csname bibfnamefont\endcsname\relax
  \def\bibfnamefont#1{#1}\fi
\expandafter\ifx\csname citenamefont\endcsname\relax
  \def\citenamefont#1{#1}\fi
\expandafter\ifx\csname url\endcsname\relax
  \def\url#1{\texttt{#1}}\fi
\expandafter\ifx\csname urlprefix\endcsname\relax\def\urlprefix{URL }\fi
\providecommand{\bibinfo}[2]{#2}
\providecommand{\eprint}[2][]{\url{#2}}

\bibitem[{\citenamefont{Zurek}(2003)}]{zurek}
\bibinfo{author}{\bibfnamefont{W.~H.} \bibnamefont{Zurek}},
  \bibinfo{journal}{Rev. Mod. Phys.} \textbf{\bibinfo{volume}{75}},
  \bibinfo{pages}{715} (\bibinfo{year}{2003}).

\bibitem[{\citenamefont{Bell}(1964)}]{bell_original}
\bibinfo{author}{\bibfnamefont{J.~S.} \bibnamefont{Bell}},
  \bibinfo{journal}{Physics} \textbf{\bibinfo{volume}{1}}, \bibinfo{pages}{195}
  (\bibinfo{year}{1964}).

\bibitem[{\citenamefont{Mermin}(1980)}]{mermin_1980}
\bibinfo{author}{\bibfnamefont{N.~D.} \bibnamefont{Mermin}},
  \bibinfo{journal}{Phys. Rev. D} \textbf{\bibinfo{volume}{22}},
  \bibinfo{pages}{356} (\bibinfo{year}{1980}).

\bibitem[{\citenamefont{Kofler and Brukner}(2007)}]{kofler}
\bibinfo{author}{\bibfnamefont{J.}~\bibnamefont{Kofler}} \bibnamefont{and}
  \bibinfo{author}{\bibfnamefont{{\v{C}}.}~\bibnamefont{Brukner}},
  \bibinfo{journal}{Phys. Rev. Lett.} \textbf{\bibinfo{volume}{99}},
  \bibinfo{pages}{180403} (\bibinfo{year}{2007}).

\bibitem[{\citenamefont{Jeong et~al.}(2014)\citenamefont{Jeong, Lim, and
  Kim}}]{kim}
\bibinfo{author}{\bibfnamefont{H.}~\bibnamefont{Jeong}},
  \bibinfo{author}{\bibfnamefont{Y.}~\bibnamefont{Lim}}, \bibnamefont{and}
  \bibinfo{author}{\bibfnamefont{M.~S.} \bibnamefont{Kim}},
  \bibinfo{journal}{Phys. Rev. Lett.} \textbf{\bibinfo{volume}{112}},
  \bibinfo{pages}{010402} (\bibinfo{year}{2014}).

\bibitem[{\citenamefont{Werner}(1989)}]{werner_1989}
\bibinfo{author}{\bibfnamefont{R.~F.} \bibnamefont{Werner}},
  \bibinfo{journal}{Phys. Rev. A} \textbf{\bibinfo{volume}{40}},
  \bibinfo{pages}{4277} (\bibinfo{year}{1989}).

\bibitem[{\citenamefont{Glauber}(1963)}]{glaub_pdist}
\bibinfo{author}{\bibfnamefont{R.~J.} \bibnamefont{Glauber}},
  \bibinfo{journal}{Phys. Rev. Lett.} \textbf{\bibinfo{volume}{131}},
  \bibinfo{pages}{2766} (\bibinfo{year}{1963}).

\bibitem[{\citenamefont{Cahill and Glauber}(1969)}]{cahill_gluaber}
\bibinfo{author}{\bibfnamefont{K.~E.} \bibnamefont{Cahill}} \bibnamefont{and}
  \bibinfo{author}{\bibfnamefont{R.~J.} \bibnamefont{Glauber}},
  \bibinfo{journal}{Phys. Rev. Lett.} \textbf{\bibinfo{volume}{177}},
  \bibinfo{pages}{1882} (\bibinfo{year}{1969}).

\bibitem[{\citenamefont{Kenfack and
  {\.Z}yczkowski}(2004)}]{kenfack2004negativity}
\bibinfo{author}{\bibfnamefont{A.}~\bibnamefont{Kenfack}} \bibnamefont{and}
  \bibinfo{author}{\bibfnamefont{K.}~\bibnamefont{{\.Z}yczkowski}},
  \bibinfo{journal}{J. Opt. B: Quantum Semiclassical Opt.}
  \textbf{\bibinfo{volume}{6}}, \bibinfo{pages}{396} (\bibinfo{year}{2004}).

\bibitem[{\citenamefont{Ferraro and Paris}(2012)}]{ferraro2012nonclassicality}
\bibinfo{author}{\bibfnamefont{A.}~\bibnamefont{Ferraro}} \bibnamefont{and}
  \bibinfo{author}{\bibfnamefont{M.~G.} \bibnamefont{Paris}},
  \bibinfo{journal}{Phys. Rev. Lett.} \textbf{\bibinfo{volume}{108}},
  \bibinfo{pages}{260403} (\bibinfo{year}{2012}).

\bibitem[{\citenamefont{Marek et~al.}(2009)\citenamefont{Marek, Kim, and
  Lee}}]{marek2009nonclassicality}
\bibinfo{author}{\bibfnamefont{P.}~\bibnamefont{Marek}},
  \bibinfo{author}{\bibfnamefont{M.}~\bibnamefont{Kim}}, \bibnamefont{and}
  \bibinfo{author}{\bibfnamefont{J.}~\bibnamefont{Lee}},
  \bibinfo{journal}{Phys. Rev. A} \textbf{\bibinfo{volume}{79}},
  \bibinfo{pages}{052315} (\bibinfo{year}{2009}).

\bibitem[{\citenamefont{Taghiabadi et~al.}(2016)\citenamefont{Taghiabadi,
  Akhtarshenas, and Sarbishaei}}]{taghiabadi2016}
\bibinfo{author}{\bibfnamefont{R.}~\bibnamefont{Taghiabadi}},
  \bibinfo{author}{\bibfnamefont{S.~J.} \bibnamefont{Akhtarshenas}},
  \bibnamefont{and}
  \bibinfo{author}{\bibfnamefont{M.}~\bibnamefont{Sarbishaei}},
  \bibinfo{journal}{QIP} \textbf{\bibinfo{volume}{15}}, \bibinfo{pages}{1999}
  (\bibinfo{year}{2016}).

\bibitem[{\citenamefont{Gerry and Knight}(2005)}]{gerry_knight}
\bibinfo{author}{\bibfnamefont{C.}~\bibnamefont{Gerry}} \bibnamefont{and}
  \bibinfo{author}{\bibfnamefont{P.}~\bibnamefont{Knight}},
  \emph{\bibinfo{title}{Introductory Quantum Optics}}
  (\bibinfo{publisher}{Cambridge University Press}, \bibinfo{year}{2005}).

\bibitem[{\citenamefont{Sekatski et~al.}(2014)\citenamefont{Sekatski, Gisin,
  and Sangouard}}]{gisin_2014}
\bibinfo{author}{\bibfnamefont{P.}~\bibnamefont{Sekatski}},
  \bibinfo{author}{\bibfnamefont{N.}~\bibnamefont{Gisin}}, \bibnamefont{and}
  \bibinfo{author}{\bibfnamefont{N.}~\bibnamefont{Sangouard}},
  \bibinfo{journal}{Phys. Rev. Lett.} \textbf{\bibinfo{volume}{113}},
  \bibinfo{pages}{090403} (\bibinfo{year}{2014}).

\bibitem[{\citenamefont{Raeisi et~al.}(2011)\citenamefont{Raeisi, Sekatski, and
  Simon}}]{raeisi_2011}
\bibinfo{author}{\bibfnamefont{S.}~\bibnamefont{Raeisi}},
  \bibinfo{author}{\bibfnamefont{P.}~\bibnamefont{Sekatski}}, \bibnamefont{and}
  \bibinfo{author}{\bibfnamefont{C.}~\bibnamefont{Simon}},
  \bibinfo{journal}{Phys. Rev. Lett.} \textbf{\bibinfo{volume}{107}},
  \bibinfo{pages}{250401} (\bibinfo{year}{2011}).

\bibitem[{\citenamefont{Ramanathan et~al.}(2011)\citenamefont{Ramanathan,
  Paterek, Kay, Kurzy\ifmmode~\acute{n}\else \'{n}\fi{}ski, and
  Kaszlikowski}}]{PhysRevLett.107.060405}
\bibinfo{author}{\bibfnamefont{R.}~\bibnamefont{Ramanathan}},
  \bibinfo{author}{\bibfnamefont{T.}~\bibnamefont{Paterek}},
  \bibinfo{author}{\bibfnamefont{A.}~\bibnamefont{Kay}},
  \bibinfo{author}{\bibfnamefont{P.}~\bibnamefont{Kurzy\ifmmode~\acute{n}\else
  \'{n}\fi{}ski}}, \bibnamefont{and}
  \bibinfo{author}{\bibfnamefont{D.}~\bibnamefont{Kaszlikowski}},
  \bibinfo{journal}{Phys. Rev. Lett.} \textbf{\bibinfo{volume}{107}},
  \bibinfo{pages}{060405} (\bibinfo{year}{2011}).

\bibitem[{\citenamefont{Leggett and Garg}(1985)}]{PhysRevLett.54.857}
\bibinfo{author}{\bibfnamefont{A.~J.} \bibnamefont{Leggett}} \bibnamefont{and}
  \bibinfo{author}{\bibfnamefont{A.}~\bibnamefont{Garg}},
  \bibinfo{journal}{Phys. Rev. Lett.} \textbf{\bibinfo{volume}{54}},
  \bibinfo{pages}{857} (\bibinfo{year}{1985}).

\bibitem[{\citenamefont{Sanders}(2012)}]{sanders2012review}
\bibinfo{author}{\bibfnamefont{B.~C.} \bibnamefont{Sanders}},
  \bibinfo{journal}{J Phys. A Math Theor.} \textbf{\bibinfo{volume}{45}},
  \bibinfo{pages}{244002} (\bibinfo{year}{2012}).

\bibitem[{\citenamefont{Jeong et~al.}(2009)\citenamefont{Jeong, Paternostro,
  and Ralph}}]{PhysRevLett.102.060403}
\bibinfo{author}{\bibfnamefont{H.}~\bibnamefont{Jeong}},
  \bibinfo{author}{\bibfnamefont{M.}~\bibnamefont{Paternostro}},
  \bibnamefont{and} \bibinfo{author}{\bibfnamefont{T.~C.} \bibnamefont{Ralph}},
  \bibinfo{journal}{Phys. Rev. Lett.} \textbf{\bibinfo{volume}{102}},
  \bibinfo{pages}{060403} (\bibinfo{year}{2009}).

\bibitem[{\citenamefont{Kofler and Brukner}(2008)}]{PhysRevLett.101.090403}
\bibinfo{author}{\bibfnamefont{J.}~\bibnamefont{Kofler}} \bibnamefont{and}
  \bibinfo{author}{\bibfnamefont{{\v{C}}.}~\bibnamefont{Brukner}},
  \bibinfo{journal}{Phys. Rev. Lett.} \textbf{\bibinfo{volume}{101}},
  \bibinfo{pages}{090403} (\bibinfo{year}{2008}).

\end{thebibliography}

\appendix 

\section{Bell quantity }\label{bell_appendix}

\subsection{Resolution coarse-graining}
The Fock state basis expansion of the odd and even coherent states is given by
\begin{equation}
\begin{split}
\ket{\alpha_e} &= C_e\sum\limits_{n = 0}^{\infty} \frac{\alpha^{2n}}{\sqrt{(2n)!}}\ket{2n} , \\
\ket{\alpha_o} &=  C_o\sum\limits_{n = 0}^{\infty} \frac{\alpha^{2n + 1}}{\sqrt{(2n + 1)!}}\ket{2n + 1},
\end{split}
\end{equation}
with, $C_e = (\cosh (|\alpha|^2))^{-1} $  and $C_o = (\sinh (|\alpha|^2))^{-1} $. For the resolution coarse-grained measurement operator, the first term in equation (\ref{bell_quanity_def}) for the Bell quantity can be written as

\begin{equation}\label{eab}
\begin{split}
E_{ab} = \frac{1}{2}\left( \langle \alpha_e | O_{\delta}(\theta_a) |\alpha_e \rangle \langle \alpha_o | O_{\delta}(\theta_b) |\alpha_o \rangle + \langle \alpha_o | O_{\delta}(\theta_a) |\alpha_o \rangle \langle \alpha_e | O_{\delta}(\theta_b) |\alpha_e \rangle  \right. \\
\left. +\langle \alpha_o | O_{\delta}(\theta_a) |\alpha_e \rangle \langle \alpha_e | O_{\delta}(\theta_b) |\alpha_o \rangle + \langle \alpha_e | O_{\delta}(\theta_a) |\alpha_o \rangle \langle \alpha_o | O_{\delta}(\theta_b) |\alpha_e \rangle   \right),
\end{split}
\end{equation}
where, $O_{\delta}(\theta_0) \equiv U^{\dagger}(\theta) O_{\delta}U(\theta_0)$. The action of $O_{\delta}$ on odd and even coherent states is as given below.

\begin{equation}
\begin{split}
O_{\delta}\ket{\alpha_e} &= \sum_{k = -\infty}^{\infty} P_{\delta}(k) \left( O_{+}^{k} - O_{-}^{k} \right) \ket{\alpha_e}, \\
&= \ket{\alpha_e} - 2C_e\sum\limits_{k = 0}^{\infty} P_{\delta}(k)\sum\limits_{n = 0}^{k}\frac{\alpha^{2n}}{\sqrt{(2n)!}}\ket{2n} , \\
&= \ket{\alpha_e} - \ket{M_e},	
\end{split}
\end{equation}
where $\ket{M_e} =  2C_e\sum\limits_{k = 0}^{\infty} P_{\delta}(k)\sum\limits_{n = 0}^{k}\frac{\alpha^{2n}}{\sqrt{(2n)!}}\ket{2n} $. Similarly,

\begin{equation}
\begin{split}
O_{\delta}\ket{\alpha_o} &=  2C_o\sum\limits_{k = 1}^{\infty} P_{\delta}(k)\sum\limits_{n = 0}^{k - 1}\frac{\alpha^{2n + 1}}{\sqrt{(2n +1)!}}\ket{2n + 1} - \ket{\alpha_o} ,  \\
&= \ket{M_o} - \ket{\alpha_e}  ,
\end{split}
\end{equation}
where $\ket{M_o} =  2C_o\sum\limits_{k = 1}^{\infty} P_{\delta}(k)\sum\limits_{n = 0}^{k - 1}\frac{\alpha^{2n + 1}}{\sqrt{(2n +1)!}}\ket{2n + 1}  $ and we have used the fact that $P_{\delta}(-k) = P_{\delta}(k) $. Using the above calculation now we can write down the terms appearing in the Bell function as

\begin{equation}
\begin{split}
\langle \alpha_e | O_{\delta}(\theta) | \alpha_e \rangle &= \langle \alpha_e | U^{\dagger}(\theta) O_{\delta} U(\theta) | \alpha_e \rangle ,\\
&= (\cos\theta \bra{\alpha_e} + \sin\theta \bra{\alpha_o}) O_{\delta}(\cos\theta \ket{\alpha_e} + \sin\theta \ket{\alpha_o}) , \\
&= (\cos\theta \bra{\alpha_e} + \sin\theta \bra{\alpha_o}) \left\lbrace \cos\theta ( \ket{\alpha_e} - \ket{M_e}) + \sin\theta ( \ket{M_o} - \ket{\alpha_o})\right\rbrace , \\
&=\cos2\theta - A \cos^2\theta  +B\sin^2\theta ,
\end{split}
\end{equation}
where $A = \braket{\alpha_e | M_e}$ and $B = \braket{\alpha_o | M_o}$. Similary we find that

\begin{equation}
\begin{split}
\langle \alpha_o | O_{\delta}(\theta) | \alpha_o \rangle &= -\cos2\theta - A \sin^2\theta  + B\cos^2\theta ,\\
\langle \alpha_e | O_{\delta}(\theta) | \alpha_o \rangle &= \cos\theta\sin\theta ( 2   - \left(A + B \right) ) = \langle \alpha_o | O_{\delta}(\theta) | \alpha_e \rangle .
\end{split}
\end{equation}
Substituting in equation (\ref{eab}) gives

\begin{equation}
\begin{split}
E_{ab} = \frac{1}{2} (\cos2\theta_a - A \cos^2\theta_a  +B\sin^2\theta_a ) (-\cos2\theta_b - A \sin^2\theta_b  + B\cos^2\theta_b ) \\
+ \frac{1}{2}(-\cos2\theta_a - A \sin^2\theta_a  + B\cos^2\theta_a ) (\cos2\theta_b - A \cos^2\theta_b  +B\sin^2\theta_b ) \\
+ \left( \cos\theta_a\sin\theta_a \sin\theta_b\cos\theta_b \right) \left( 2 - (A + B) \right)^2 .
\end{split}
\end{equation}
After some simplification, this becomes

\begin{equation}
E_{ab} =  \cos 2(\theta_a + \theta_b) \left( -1 + A + B -\frac{1}{4}(A + B)^2 \right) + \frac{1}{2}(A - B)^2.
\end{equation}
The maximum value of the Bell observable from equation (\ref{bell_quanity_def}) under resolution coarse-graining is then given by
\begin{equation}
\begin{split}
  B_{\delta} = \max\limits_{\theta_a, \theta_b, \theta_c, \theta_d} \mathcal{F}(\theta_a, \theta_b, \theta_c, \theta_d) \left( -1 + A + B - \frac{1}{4} (A + B)^2 \right)  + \frac{1}{2}(A - B)^2,
\end{split}
\end{equation}  
where
\begin{equation}
\begin{split}
\mathcal{F}(\theta_a, \theta_b, \theta_c, \theta_d) = \cos\left(2\theta_a + 2\theta_b \right) + \cos\left(2\theta_c + 2\theta_b \right)   + \cos\left(2\theta_a + 2\theta_d \right) - \cos\left(2\theta_c + 2\theta_d \right),
\end{split}
\end{equation}
with $\max\limits_{\theta_a, \theta_b, \theta_c, \theta_d} \mathcal{F}(\theta_a, \theta_b, \theta_c, \theta_d) = 2\sqrt{2}$. 
Note that $A$ and $B$ can be written as
\begin{equation}
\begin{split}
A  &= 2C_e^2\sum\limits_{k = 0}^{\infty} \sum\limits_{n = 0}^{k} P_{\delta}(k)\frac{(|\alpha|^2)^{2n}}{(2n)!} ,\\
B  &= 2C_o^2 \sum\limits_{k = 1}^{ \infty} \sum\limits_{n = 0}^{k - 1} P_{\delta}(k)\frac{(|\alpha|^2)^{2n + 1}}{(2n + 1)!} .\\
\end{split}
\end{equation}

\subsection{Reference coarse-graining}
For the reference coarse-grained measurement operator, the first term in equation (\ref{bell_quanity_def}) for the Bell quantity can be expanded as
\begin{equation}\label{eabref}
\begin{split}
E_{ab} = \frac{1}{2}\left( \langle \alpha_e | O_{\Delta}(\theta_a) |\alpha_e \rangle \langle \alpha_o | O_{\Delta}(\theta_b) |\alpha_o \rangle + \langle \alpha_o | O_{\Delta}(\theta_a) |\alpha_o \rangle \langle \alpha_e | O_{\Delta}(\theta_b) |\alpha_e \rangle  \right. \\
\left. +\langle \alpha_o | O_{\Delta}(\theta_a) |\alpha_e \rangle \langle \alpha_e | O_{\Delta}(\theta_b) |\alpha_o \rangle + \langle \alpha_e | O_{\Delta}(\theta_a) |\alpha_o \rangle \langle \alpha_o | O_{\Delta}(\theta_b) |\alpha_e \rangle   \right).
\end{split}
\end{equation}
The individual terms can be computed as
\begin{align}
\langle \alpha_e |  O_{\Delta}(\theta_0) |\alpha_e \rangle =& \langle \alpha_e | \left( \int\limits_{\infty}^{\infty}  \diff \theta  P_{\Delta} (\theta - \theta_0)U_{\dagger} (\theta) O^{0}U(\theta) \right)|\alpha_e \rangle  ,\\
&= \int\limits_{\infty}^{\infty} \diff\theta  P_{\Delta}(\theta -\theta_0 )  (\cos\theta \bra{\alpha_e} + \sin\theta \bra{\alpha_o}) O^0(\cos\theta \ket{\alpha_e} + \sin\theta \ket{\alpha_o}) ,  \\
&= \int\limits_{\infty}^{\infty} \diff\theta  \frac{1}{\sqrt{2\pi}\Delta}\exp(-\frac{(\theta - \theta_0)^2}{2\Delta^2})   \cos2 \theta  , \\
&= e^{-2\Delta^2}\cos 2 \theta_0.
\end{align}
Similarly,

\begin{align}
\langle \alpha_o |  O_{\Delta}(\theta_0) |\alpha_o \rangle &= -e^{-2\Delta^2}\cos 2 \theta_0 ,\\
\langle \alpha_e |  O_{\Delta}(\theta_0) |\alpha_o \rangle &= e^{-2\Delta^2}\sin 2 \theta_0 = \langle \alpha_o |  O_{\Delta}(\theta_0) |\alpha_e \rangle.
\end{align}
Substituting the above results in eqn (\ref{eabref}) gives
\begin{align*}
E_{ab} &= \frac{e^{-4\Delta^2} }{2} \left( -\cos 2\theta_a \cos 2\theta_b - \cos 2\theta_a \cos 2\theta_b  + \sin 2\theta_a \sin 2\theta_b + \sin 2\theta_a \sin 2\theta_b  \right) ,\\
&= -e^{-4\Delta^2}\cos(2\theta_a + 2\theta_b).
\end{align*}
Using equation (\ref{bell_quanity_def}) gives the Bell quantity to be
\begin{equation}
B_{\Delta}(\theta_a, \theta_b, \theta_c, \theta_d) = \mathcal{F}(\theta_a, \theta_b, \theta_c, \theta_d)e^{-4\Delta^2},
\end{equation}
where 
\begin{equation}
\mathcal{F}(\theta_a, \theta_b, \theta_c, \theta_d) = -\left( \cos(2\theta_a + 2\theta_b) + \cos(2\theta_c + 2\theta_b) + \cos(2\theta_a + 2\theta_d) - \cos(2\theta_c + 2\theta_d) \right) .
\end{equation}
Maximizing the Bell quantity over all values of $\theta_i$ gives

\begin{equation}
\max_{\theta_a, \theta_b, \theta_c, \theta_d}B_{\Delta}(\theta_a, \theta_b, \theta_c, \theta_d) = B_{\Delta} = 2\sqrt{2} e^{-4\Delta^2}.
\end{equation}

\section{Post-measurement state for single mode Fock state} \label{fock_appendix}

The initial state is chosen to be a even Fock state $| 2n \rangle \langle 2n |$. The unitary operator is chosen to be a rotation between even and odd Fock states as given below.

\begin{equation}\label{unitoddeven}
\begin{split}
&U(\theta) \ket{2n} = \cos \theta \ket{2n} + \sin\theta \ket{2n - 1} ,\\
&U(\theta) \ket{2n -1} = \sin \theta \ket{2n} - \cos\theta \ket{2n - 1} .
\end{split}
\end{equation}

\subsection{Resolution coarse-graining}
 For resolution coarse-graining, the post-measurement state will be the same as the initial state. 
\begin{equation}
\begin{split}
\rho_{res}(\delta) &=  \sum\limits_{k = -\infty}^{\infty} P_{\delta}(k) \left(  O_{+}^{k} | 2n\rangle\langle 2n | O_{+}^{k}  + O_{-}^{k} | 2n\rangle\langle 2n | O_{-}^{k}  \right), \\
&= | 2n \rangle\langle 2n |.
\end{split}
 \end{equation} 
The P-distribution of a Fock state $\ket{n}$ can be calculated as \citep{gerry_knight}

\begin{equation}\label{mehta_form}
\begin{split}
P(\gamma) &= \frac{1}{\pi^2}\int \Tr \left( \ketbra{n}{n} e^{\lambda \hat{a}^{\dagger}} e^{-\lambda^* \hat{a}}  \right) e^{\lambda^*\gamma - \lambda \gamma^*} \Diff2 \lambda ,\\
&= \frac{1}{\pi^2} \sum\limits_{m = 0}^{n}\frac{1}{m!} {n \choose m}\int (-|\lambda|^2)^n e^{\lambda^* \gamma - \lambda\gamma^*}\Diff2 \lambda, \\
&= \frac{1}{\pi^2} \sum\limits_{m = 0}^{n}\frac{1}{m!} {n \choose m} \left( \frac{\partial}{\partial \gamma} \frac{\partial}{\partial \gamma^*} \right)^m \int  e^{\lambda^* \gamma - \lambda\gamma^*}\Diff2 \lambda, \\ 
&=  \sum\limits_{m = 0}^{n}\frac{1}{m!} {n \choose m} \left( \nabla_{\gamma}^2 \right)^m  \delta^{(2)}(\gamma) , \\
&= L_{n}(-\nabla_{\gamma}^2)  \delta^{(2)}(\gamma).
\end{split}
\end{equation}

\subsection{Reference Coarse-graining}
For reference coarse-graining, the post-measurement state is
\begin{equation}
\begin{split}
 \rho_{ref}(\Delta) = \int P_{\Delta}(\theta - \theta_a) \left(  O^{0}_{+}(\theta) |2n \rangle\langle 2n |O_{+}^{0}(\theta) \right.  
 \left. + O^{0}_{-}(\theta) |2n \rangle\langle 2n |O_{-}^{0}(\theta) \right) \diff\theta  ,
 \end{split}
 \end{equation} 
Note that
\begin{equation}
\begin{split}
O^{0}_{+}(\theta) |2n \rangle\langle 2n |O_{+}^{0}(\theta)  
 &= \cos^2\theta \left(  \cos^2\theta |2n\rangle \langle 2n |  + \sin^2\theta |2n + 1\rangle \langle 2n + 1 | \right. \\
 & \hspace{3cm}+  \left. \sin\theta\cos\theta \left\lbrace |2n\rangle \langle 2n + 1 | + |2n + 1\rangle \langle 2n | \right\rbrace \right),
 \end{split}
\end{equation}
and
\begin{equation}
\begin{split}
O^{0}_{-}(\theta) |2n \rangle\langle 2n |O_{-}^{0}(\theta)  
 &= \sin^2\theta \left(  \sin^2\theta |2n\rangle \langle 2n |  + \cos^2\theta |2n - 1\rangle \langle 2n + 1 | \right. \\
 & \hspace{3cm} -  \left. \sin\theta\cos\theta \left\lbrace |2n\rangle \langle 2n + 1 | + |2n + 1\rangle \langle 2n | \right\rbrace \right).
\end{split}.
\end{equation}
Identifying $\rho (\theta)$ as $O^{0}_{+}(\theta) |2n \rangle\langle 2n |O_{+}^{0}(\theta) + O^{0}_{-}(\theta) |2n \rangle\langle 2n |O_{-}^{0}(\theta)$, and using the above result,

\begin{equation}
\begin{split}
\rho(\theta) &= (\cos^4 \theta + \sin^4\theta) | 2n \rangle \langle 2n | + (\sin\theta \cos\theta(\cos^2 \theta - \sin^2\theta) | 2n \rangle \langle 2n + 1| \\ 
& \hspace{2cm} +(\sin\theta \cos\theta(\cos^2 \theta - \sin^2\theta) | 2n + 1 \rangle \langle 2n | + (2\cos^2 \theta \sin^2\theta) | 2n + 1 \rangle \langle 2n + 1| ,
\end{split}
\end{equation}
or expressed in the basis $\lbrace \ket{2n}, \ket{2n + 1}$,
\begin{equation}
\rho (\theta) = \frac{1}{4}\left(
\begin{array}{cc} 
3 + \cos 4\theta & \sin 4\theta \\
\sin 4\theta &1 - \cos 4\theta
\end{array}
 \right).
\end{equation}
The post-measurement state 
\begin{align*}
 \rho_{ref}(\Delta) &= \int d\theta P_{\Delta}(\theta - \theta_a) \rho (\theta), \\
 &=  \frac{1}{4}\left(
\begin{array}{cc} 
3 + e^{-8\Delta^2}\cos 4\theta_a & e^{-8\Delta^2}\sin 4\theta_a \\
e^{-8\Delta^2}\sin 4\theta_a &1 - e^{-8\Delta^2}\cos 4\theta_a
\end{array}
 \right).
\end{align*}
The P-distribution can be calculated using the same method as in equation (\ref{mehta_form}). Let us now calculate for each matrix element individually. Let us define
\begin{equation}
P_{n, m} (\gamma) = \frac{1}{\pi^2}\int \Tr \left( \ketbra{n}{m} e^{\lambda \hat{a}^{\dagger}} e^{\lambda^* \hat{a}}  \right) e^{\lambda^* \gamma - \lambda\gamma^*}\Diff2 \lambda .
\end{equation}

\noindent $P_{n,n}(\gamma)$ and $P_{n+1, n+1}(\gamma)$ are already known from equation (\ref{mehta_form}). To compute $P_{n+1, n}(\gamma)$ and $P_{n, n+1}(\gamma)$, consider the traces given below.
\begin{equation}
\begin{split}
\Tr \left( \ketbra{n + 1}{n} e^{\lambda \hat{a}^{\dagger}} e^{-\lambda^* \hat{a}}  \right) &=  \braket{n |  e^{\lambda \hat{a}^{\dagger}} e^{-\lambda^* \hat{a}}     | n + 1} ,\\
&= \sum\limits_{m = 0}^{n} \sum\limits_{p = 0}^{n + 1} \left( \frac{\lambda^m}{m!}\sqrt{\frac{n!}{(n - m)!}}\bra{n - m} \right) \left( \frac{(-\lambda^*)^p}{p!}\sqrt{\frac{(n + 1)!}{(n + 1 - p)!}}\ket{n + 1 - p} \right) ,\\
&= \sum\limits_{m = 0}^{n} \frac{-\lambda^* \sqrt{n + 1}}{ (m + 1)!} {n \choose m} (- |\lambda|^2)^m  .
\end{split}
\end{equation}
Similarly,
\begin{equation}
\begin{split}
\Tr \left( \ketbra{n }{n + 1} e^{\lambda \hat{a}^{\dagger}} e^{-\lambda^* \hat{a}}  \right) &= \sum\limits_{m = 0}^{n} \frac{\lambda \sqrt{n + 1}}{(m + 1)!} {n  \choose m} (- |\lambda|^2)^m  .
\end{split}
\end{equation}
Proceeding as before,
\begin{align*}
P_{n + 1, n}(\gamma) &=   \frac{1}{\pi^2}\sum\limits_{m = 0}^{n} \frac{ \sqrt{n + 1}}{(m + 1)!} {n \choose m} \int  (- |\lambda|^2)^m (-\lambda^*) e^{\lambda^* \gamma - \lambda\gamma^*}\Diff2 \lambda, \\
&=    \frac{1}{\pi^2}\sum\limits_{m = 0}^{n } \frac{ \sqrt{n+ 1}}{(m + 1)!} {n \choose m} \left( -\frac{\partial}{\partial \gamma} \right) \left( \frac{\partial}{\partial \gamma} \frac{\partial}{\partial \gamma^*}  \right)^m  \int   e^{\lambda^* \gamma - \lambda\gamma^*}\Diff2 \lambda, \\
& = \left( -\frac{\partial}{\partial \gamma} \right) M_{n}(\nabla_{\gamma}^{2})\delta^{(2)}(\gamma),
\end{align*}
where
\begin{equation}
M_{n}(x) \equiv \sum\limits_{m = 0}^{n} \frac{ \sqrt{n + 1}}{(m + 1)!} {n \choose m} x^m,
\end{equation}
and
\begin{align*}
P_{n + 1 , n}(\gamma) &= \left( -\frac{\partial}{\partial \gamma^*} \right) M_{n}(\nabla_{\gamma}^{2})\delta^{(2)}(\gamma).
\end{align*}
Putting these results together, the full P-distribution for the post-measurement state is
\begin{equation}
\begin{split}
\mathcal{P}(\gamma) = \frac{1}{4} \left\lbrace (3 + e^{-8\Delta^2}\cos 4\theta_a) L_{2n + 1}(-\nabla_{\gamma}^2)  - e^{-8\Delta^2}\sin 4\theta_a \left ( \frac{\partial}{\partial\gamma} + \frac{\partial}{\partial\gamma^*} \right )M_{2n}(\nabla_{\gamma}^2) \right. \\ \left.
+ (1  - e^{-8\Delta^2}\cos 4\theta_a )L_{2n - 1}(-\nabla_{\gamma}^2) \right\rbrace  .
\end{split}
\end{equation}

\section{Post-measurement sate for two-mode cat state} \label{post_meas_cat_appendix}
The initial state is chosen to be:
\begin{equation}
\ket{\psi_{in}} = \frac{1}{\sqrt{2}}\left( \ket{\alpha_e}\ket{\alpha_o} + \ket{\alpha_o}\ket{\alpha_e} \right) .
\end{equation}
The corresponding density matrix is
\begin{align*}
\rho = | \psi_{in} \rangle\langle \psi_{in} | = \frac{1}{2} \left(  |\alpha_e \rangle \langle \alpha_e| \otimes |\alpha_o \rangle \langle \alpha_o| + |\alpha_e \rangle \langle \alpha_o| \otimes |\alpha_o \rangle \langle \alpha_e| + |\alpha_o \rangle \langle \alpha_e| \otimes |\alpha_e \rangle \langle \alpha_o| + |\alpha_o \rangle \langle \alpha_o| \otimes |\alpha_e \rangle \langle \alpha_e|   \right).
\end{align*}
\subsubsection{Reference Coarse-graining}

The measurement operator is
\begin{equation}
\begin{split}
O_{\Delta}(\theta_a, \theta_b) = \int d\theta_1 \int d\theta_2 P_{\Delta}(\theta_1 - \theta_a)P_{\Delta}(\theta_2 - \theta_b)   \left[ O^{0}(\theta_1) \otimes O^{0}(\theta_2) \right],
\end{split}
\end{equation}
where $O^{0}(\theta) \equiv U^{\dagger}(\theta) O^{0} U(\theta) $ and the action of the unitary is the same as in equation (\ref{rot_alpha_def}). The post-measurement state is then given by

\begin{equation}\label{ref_post_meas_appendix_def}
\begin{split}
\rho_{ref} = \int d\theta_1 \int d\theta_2 P_{\Delta}(\theta_1 - \theta_a)P_{\Delta}(\theta_2 - \theta_b) \left\lbrace \phantom{\frac{1}{2}} \hspace{-0.3cm}
O_{+}(\theta_1) \otimes O_{+}(\theta_2)] \rho [O_{+}(\theta_1) \otimes O_{+}(\theta_2)]  \right.\\ \left.
+ [O_{+}(\theta_1) \otimes O_{-}(\theta_2)] \rho [O_{+}(\theta_1) \otimes O_{-}(\theta_2)]  \right.\\ \left.
+ [O_{-}(\theta_1) \otimes O_{+}(\theta_2)] \rho [O_{-}(\theta_1) \otimes O_{+}(\theta_2)]\right.\\ \left.
+ [O_{-}(\theta_1) \otimes O_{-}(\theta_2)] \rho [O_{-}(\theta_1) \otimes O_{-}(\theta_2)]
\phantom{\frac{1}{2}} \hspace{-.3cm}\right\rbrace.
\end{split}
\end{equation}
Calculating term by term, 

\begin{equation}
\begin{split}
O_{+}(\theta_1) \otimes O_{+}(\theta_2)] \rho [O_{+}(\theta_1) \otimes O_{+}(\theta_2)] = \frac{1}{2} \left\lbrace \phantom{\frac{1}{2}} \hspace{-0.3cm} \cos^2\theta_1 \alketbra{e}{e}_{\theta_1}  \otimes \sin^2\theta_2 \alketbra{e}{e}_{\theta_2}  \right. \\ \left.
\cos\theta_1\sin\theta_1 \alketbra{e}{e}_{\theta_1}  \otimes \cos\theta_2\sin\theta_2 \alketbra{e}{e}_{\theta_2}  \right. \\ \left.
\cos\theta_1\sin\theta_1 \alketbra{e}{e}_{\theta_1}  \otimes \cos\theta_2\sin\theta_2 \alketbra{e}{e}_{\theta_2}  \right. \\ \left.
sin^2\theta_1 \alketbra{e}{e}_{\theta_1}  \otimes \cos^2\theta_2 \alketbra{e}{e}_{\theta_2}   \phantom{\frac{1}{2}} \hspace{-.3cm} \right\rbrace, \\
= \frac{1}{2}\sin^2 (\theta_1 + \theta_2) \alketbra{e}{e}_{\theta_1}  \otimes \alketbra{e}{e}_{\theta_2} ,
\end{split}
\end{equation}
where $\ketbra{*}{*}_{\theta} \equiv U^{\dagger}(\theta)\ketbra{*}{*}U(\theta) $. Similarly, 

\begin{align}
O_{+}(\theta_1) \otimes O_{-}(\theta_2)] \rho [O_{+}(\theta_1) \otimes O_{-}(\theta_2)] &= \frac{1}{2}\cos^2 (\theta_1 + \theta_2) \alketbra{e}{e}_{\theta_1}  \otimes \alketbra{o}{o}_{\theta_2},  \\
O_{-}(\theta_1) \otimes O_{+}(\theta_2)] \rho [O_{-}(\theta_1) \otimes O_{+}(\theta_2)]  &= \frac{1}{2}\cos^2 (\theta_1 + \theta_2) \alketbra{o}{o}_{\theta_1}  \otimes \alketbra{e}{e}_{\theta_2}, \\
O_{-}(\theta_1) \otimes O_{-}(\theta_2)] \rho [O_{-}(\theta_1) \otimes O_{-}(\theta_2)] &=  \frac{1}{2}\sin^2 (\theta_1 + \theta_2) \alketbra{o}{o}_{\theta_1}  \otimes \alketbra{o}{o}_{\theta_2} .
\end{align}
Substituting the above results in equation (\ref{ref_post_meas_appendix_def}) gives the post-measurement state as
\begin{equation}
\begin{split}
\rho_{ref} = \int d\theta_1 \int d\theta_2 P_{\Delta}(\theta_1 - \theta_a)P_{\Delta}(\theta_2 - \theta_b) \left\lbrace \phantom{\frac{1}{2}} \hspace{-0.3cm}
\frac{1}{2}\sin^2 (\theta_1 + \theta_2) \alketbra{e}{e}_{\theta_1}  \otimes \alketbra{e}{e}_{\theta_2}   \right.\\ \left.
+ \frac{1}{2}\cos^2 (\theta_1 + \theta_2) \alketbra{e}{e}_{\theta_1}  \otimes \alketbra{o}{o}_{\theta_2}   \right.\\ \left.
+ \frac{1}{2}\cos^2 (\theta_1 + \theta_2) \alketbra{o}{o}_{\theta_1}  \otimes \alketbra{e}{e}_{\theta_2}  \right.\\ \left.
+ \frac{1}{2}\sin^2 (\theta_1 + \theta_2) \alketbra{o}{o}_{\theta_1}  \otimes \alketbra{o}{o}_{\theta_2} 
\phantom{\frac{1}{2}} \hspace{-.3cm}\right\rbrace.
\end{split}
\end{equation}
Rewriting in the unrotated basis and carrying out the integral gives
\begin{equation}\label{post_appendix}
\rho_{ref} = \left(
\begin{array}{cccc}
a &\phantom{-}b &\phantom{-}c &\phantom{-}d  \\
b &\frac{1}{2}-a &\phantom{-}d &-c \\
c &\phantom{-}d &\frac{1}{2} -a &-b \\
d &-c &-b &\phantom{-}a
\end{array} \right ),
\end{equation}
with
\begin{align}
\begin{split}
\label{a_appendix}
a = \frac{1}{16} \left( 3 - e^{-8\Delta^2}\left\lbrace\cos (4\theta_a) + \cos \left(4\theta_b\right)\right\rbrace \right.  
 \left. 
- e^{-16\Delta^2}\cos (4\theta_a + 4\theta_b)\right) 
\end{split} ,\\
\begin{split}
b = \frac{1}{16} \left( e^{-8\Delta^2}\left\lbrace\sin (4\theta_a) - \sin \left(4\theta_b\right)\right\rbrace \right. 
 \left. - e^{-16\Delta^2}\sin (4\theta_a + 4\theta_b)\right) \\
\end{split} ,\\
\begin{split}
c  = \frac{1}{16} \left( e^{-8\Delta^2}\left\lbrace - \sin (4\theta_a) + \sin \left(4\theta_b\right)\right\rbrace \right. 
 \left. 
- e^{-16\Delta^2}\sin (4\theta_a + 4\theta_b)\right)
\end{split} ,\\
\begin{split}
\label{d_appendix}
d = \frac{1}{16} \left(1 -  e^{-8\Delta^2}\left\lbrace\cos (4\theta_a) + \cos \left(4\theta_b\right)\right\rbrace \right. 
 \left.
+ e^{-16\Delta^2}\cos (4\theta_a + 4\theta_b)\right).
\end{split}
\end{align}

\subsubsection{P distribution}
The single mode P distribution calculation in equation (\ref{mehta_form}) can be extended to a two-mode calculation for the post-measurement state $\rho_{ref}$ as
\begin{equation}\label{p_dist_def_appendix_eq}
\mathcal{P}_{ref}(\beta, \gamma) = \frac{1}{\pi^4} \int \int \Tr \left( \rho_{ref} \left[ e^{\lambda \hat{a}^{\dagger}}e^{-\lambda^*\hat{a}} \otimes  e^{\tau \hat{b}^{\dagger}}e^{-\tau^*\hat{b}} \right] \right)e^{\lambda^*\beta - \lambda\beta^*} e^{\tau^*\gamma - \tau\gamma^*}\Diff2 \lambda \Diff2 \tau.
\end{equation}
Using the Schmidt decomposition for a the density matrix, we can write
\begin{equation}
\rho_{ref} =  \sum\limits_{i, j, k, l}\rho_{i,j,k,l}\ketbra{i}{j} \otimes \ketbra{k}{l}.
\end{equation}
The double integral in equation \eqref{p_dist_def_appendix_eq} can be written as a sum of products of integrals given by 
\begin{equation}\label{pprod}
\mathcal{P}_{ref}(\beta, \gamma) =  \frac{1}{\pi^4}\sum\limits_{i, j, k, l}\rho_{i,j,k,l} \left ( \int \Tr \left( \ketbra{i}{j}e^{\lambda \hat{a}^{\dagger}} e^{-\lambda^*\hat{a}}\right)e^{\lambda^*\beta - \lambda\beta^*} \Diff2\lambda \right) \left ( \int \Tr \left( \ketbra{k}{l} e^{\tau \hat{b}^{\dagger}} e^{-\tau^*\hat{b}} \right) e^{\tau^*\gamma - \tau\gamma^*}\Diff2\tau \right) .
\end{equation}
Choosing the two basis for the Schmidt decomposition to be $\left\lbrace  \ket{e}, \ket{o} \right\rbrace$, with $\ket{e} \equiv \ket{\alpha_e}$ and $\ket{o} \equiv \ket{\alpha_o}$, (this is sufficient since the density matrix has support only in this subspace) will make the coefficients
\begin{equation}
  \rho_{i,j,k,l} = \Tr\left( \rho_{ref} \ketbra{i}{j} \otimes \ketbra{k}{l} \right) \hspace{2cm} i,j,k,l \in \{ e, o\},
 \end{equation} 
 which are essentially the matrix coefficients in equation (\ref{post}). So the P-distribution can be written as
\begin{equation} \label{pdist_appendix}
\mathcal{P}_{ref}(\beta, \gamma) =  \sum\limits_{i, j, k, l, \in \left\lbrace e, o\right\rbrace}\rho_{i,j,k,l}P_{ij}(\beta)P_{kl}(\gamma),
\end{equation}
where
\begin{equation}\label{pij_def}
P_{ij} (\beta)=  \frac{1}{\pi^2}\int \Tr \left( \ketbra{i}{j}e^{\lambda \hat{a}^{\dagger}} e^{-\lambda^*\hat{a}} \right)e^{\lambda^*\beta - \lambda\beta^*} \Diff2\lambda .
\end{equation}
Note that
\begin{align}
\Tr \left( \alketbra{e}{e}e^{\lambda^*\hat{a}}e^{-\lambda \hat{a}^{\dagger}}\right) &=  N_e^2 \left[ e^{\lambda\alpha^* - \lambda^*\alpha} + e^{-\lambda\alpha^* + \lambda^*\alpha} + e^{-2|\alpha|^2} (e^{-\lambda\alpha^* - \lambda^*\alpha} + e^{\lambda\alpha^* + \lambda^*\alpha} )  \right].
\end{align}
Substituting this in equation (\ref{pij_def}) gives

\begin{align}\label{preftr}
P_{ee}(\beta) = \frac{N_{e}^{2}}{\pi^2}\int \left(e^{\lambda^*(\beta - \alpha) - \lambda(\beta^* - \alpha^*)} + e^{\lambda^*(\beta + \alpha) - \lambda(\beta^* + \alpha^*)} + e^{-2|\alpha|^2} \left( e^{\lambda^*(\beta - \alpha) - \lambda(\beta^* + \alpha^*)} + e^{\lambda^*(\beta + \alpha) - \lambda(\beta^* - \alpha^*)} \right)\right) \Diff2\lambda .
\end{align}
To write this in a closed form, consider the integral
\begin{align*}
\int e^{\lambda^*(\beta - \alpha) - \lambda(\beta^* + \alpha^*)} \Diff2 \lambda &= \int e^{-2\lambda^*\alpha} e^{\lambda^*(\beta + \alpha) - \lambda(\beta^* + \alpha^*)} \Diff2 \lambda , \\
&= \int \sum\limits_{n = 0}^{\infty} \frac{(-2\alpha)^n}{n!} \left( \frac{\partial}{\partial \alpha} \right)^n e^{\lambda^*(\beta + \alpha) - \lambda(\beta^* + \alpha^*)} \Diff2 \lambda ,\\
&= \int \hat{A}(\alpha) e^{\lambda^*(\beta + \alpha) - \lambda(\beta^* + \alpha^*)} \Diff2 \lambda ,\\  
&= \hat{A}(\alpha)\int  e^{\lambda^*(\beta + \alpha) - \lambda(\beta^* + \alpha^*)} \Diff2 \lambda  ,\\
&= \hat{A} (\alpha) \pi^2 \delta^{(2)}(\alpha + \beta),
\end{align*}
where $\hat{A} \equiv \sum\limits_{n = 0}^{\infty} \dfrac{(-2\alpha)^n}{n!} \left( \dfrac{\partial}{\partial \alpha} \right)^n $. Using the fact that $\hat{A}(\alpha) = \hat{A}(-\alpha)$, the other integral can be written down as

\begin{align*}
\int e^{\lambda^*(\beta + \alpha) - \lambda(\beta^* - \alpha^*)} \Diff2 \lambda = \hat{A}  (\alpha) \pi^2 \delta^{(2)}(\alpha - \beta) .
\end{align*}

Substituting these results in equation (\ref{preftr}) gives
\begin{equation}
\begin{split}
P_{ee}(\beta) &= N_{e}^{2} \left( \delta^{(2)}(\alpha - \beta)  + \delta^{(2)}(\alpha + \beta)  + e^{-2|\alpha|^2}\hat{A}(\alpha) \left( \delta^{(2)}(\alpha - \beta)  + \delta^{(2)}(\alpha + \beta) \right) \right) ,\\
& = N_e^2   \left\lbrace 1 + e^{-2|\alpha|^2} \hat{A}(\alpha)   \right\rbrace  
 \left[ \delta^{(2)}(\alpha - \beta)  +\delta^{(2)}(\alpha + \beta)\right]. 
\end{split}
\end{equation}
Similarly, 
\begin{align}
\begin{split}
P_{eo}(\beta) = N_eN_o   \left\lbrace 1 + e^{-2|\alpha|^2} \hat{A}(\alpha)   \right\rbrace  
 \left[ \delta^{(2)}(\alpha - \beta)  - \delta^{(2)}(\alpha + \beta)\right],
\end{split} \\
\begin{split}
P_{oe}(\beta) = N_eN_o   \left\lbrace 1 - e^{-2|\alpha|^2} \hat{A}(\alpha)   \right\rbrace 
 \left[ \delta^{(2)}(\alpha - \beta)  - \delta^{(2)}(\alpha + \beta)\right],  
\end{split} \\
\begin{split}
P_{oo}(\beta) = N_o^2   \left\lbrace 1 - e^{-2|\alpha|^2} \hat{A}(\alpha)   \right\rbrace  
 \left[ \delta^{(2)}(\alpha - \beta)  + \delta^{(2)}(\alpha + \beta)\right].
\end{split}
\end{align}

\subsubsection{Wigner Function}
As we did for the P-distribution in the previous section, the Wigner function can be written as
\begin{equation}\label{wprod}
\mathcal{W}_{ref}(\beta, \gamma) =  \frac{1}{\pi^4}\sum\limits_{i, j, k, l}\rho_{i,j,k,l} W_{ij}(\beta) W_{kl}(\gamma) ,
\end{equation}
where
\begin{equation}\label{wij}
W_{ij}(\beta) =  \int \Tr \left( \ketbra{i}{j}e^{\lambda \hat{a}^{\dagger}} e^{-\lambda^*\hat{a}}\right)e^{-\frac{|\lambda|^2}{2}}e^{\lambda^*\beta - \lambda\beta^*} \Diff2\lambda .
\end{equation}
Computing the trace and carrying out the integral gives
\begin{equation}
\begin{split}
&W_{ee}(\beta) = \frac{|N_e|^2}{\pi^2} \int e^{\lambda^{*}\beta - \lambda\beta^{*}} e^{-\frac{|\lambda |^2}{2}}\left(e^{\lambda \alpha^* -\lambda^{*} \alpha} + e^{-\lambda \alpha^* + \lambda^{*} \alpha}  +e^{-2|\alpha|^2} ( e^{-\lambda \alpha^* -\lambda^{*} \alpha} + e^{\lambda \alpha^* + \lambda^{*} \alpha})\right) d^2\lambda , \\
&= \frac{2|N_e|^2}{\pi} \left( e^{-2| \alpha - \beta |^2} + e^{-2| \alpha + \beta |^2} +e^{-2|\alpha |^2} \left( e^{2(\alpha - \beta)(\alpha^* + \beta^*)} + e^{2(\alpha + \beta)(\alpha^* - \beta^*)} \right) \right) .
\end{split}
\end{equation}
Similarly,
\begin{align}
\begin{split}
W_{eo}(\beta) = \frac{2}{\pi}N_{e}N_o
\left\lbrace e^{-2 | \alpha - \beta |^2} - e^{-2 | \alpha + \beta |^2}    \right. 
\left.
+ e^{-2|\alpha|^2} \left[ e^{2(\alpha - \beta)(\alpha^* + \beta^* )}  - e^{2(\alpha + \beta)(\alpha^* - \beta^* )} \right] \right\rbrace ,
\end{split} \\
\begin{split}
W_{oe}(\beta) = \frac{2}{\pi}N_{e}N_o
\left\lbrace e^{-2 | \alpha - \beta |^2} - e^{-2 | \alpha + \beta |^2}    \right. 
\left.
- e^{-2|\alpha|^2} \left[ e^{2(\alpha - \beta)(\alpha^* + \beta^* )}  - e^{2(\alpha + \beta)(\alpha^* - \beta^* )} \right] \right\rbrace ,
\end{split} \\
\begin{split}
W_{oo}(\beta) = \frac{2}{\pi} N_o^2
\left\lbrace e^{-2 | \alpha - \beta |^2} + e^{-2 | \alpha + \beta |^2}    \right. 
\left.
- e^{-2|\alpha|^2} \left[ e^{2(\alpha - \beta)(\alpha^* + \beta^* )}  + e^{2(\alpha + \beta)(\alpha^* - \beta^* )} \right] \right\rbrace .
\end{split} 
\end{align}

\section{NOON states} \label{ent_fock_appendix}

For the NOON state:
\begin{equation}
\ket{\psi_{in}} =  \frac{1}{\sqrt{2}}\left( \ket{0}\ket{N} +  \ket{N}\ket{0}   \right),
\end{equation}
the same calculation in the Appendix \ref{post_meas_cat_appendix} leads to a post-measurement state
\begin{equation}
\rho_{ref} = \left(
\begin{array}{cccc}
a &\phantom{-}b &\phantom{-}c &\phantom{-}d  \\
b &\frac{1}{2}-a &\phantom{-}d &-c \\
c &\phantom{-}d &\frac{1}{2} -a &-b \\
d &-c &-b &\phantom{-}a
\end{array} \right ),
\end{equation}
with $a$, $b$, $c$, $d$ given by equations \eqref{a_appendix}-\eqref{d_appendix}. However, now the basis in which $\rho_{ref}$ is expressed is, $\lbrace \ket{00}, \ket{0N}, \ket{N0}, \ket{NN} \rbrace$. The Wigner function can be expressed as
\begin{equation}
\mathcal{W}_{ref}(\beta, \gamma) =  \frac{1}{\pi^4}\sum\limits_{i, j, k, l \in \lbrace e, o \rbrace}\rho_{i,j,k,l } W_{ij}(\beta) W_{kl}(\gamma) ,
\end{equation}
where now we choose $\ket{e} \equiv \ket{0}$ and $\ket{o} \equiv \ket{N} $. Note that $W_{ij}(\beta)$ is the Wigner function of the operator $\ketbra{i}{j}$, which can can be computed to be
\begin{align}
\begin{split}
W_{ee}(\beta) = \frac{2}{\pi} e^{- 2 |\beta|^2} ,
\end{split} \\
\begin{split}
W_{eo}(\beta) = \frac{2}{\pi} \frac{(2 \beta)^{N}}{\sqrt{N !}}e^{- 2 |\beta |^2},
\end{split} \\
\begin{split}
W_{oe}(\beta) = \frac{2}{\pi} \frac{(2 \beta^*)^{N}}{\sqrt{N!}}e^{- 2 |\beta|^2},
\end{split} \\
\begin{split}
W_{oo}(\beta) = -\frac{2}{\pi} L_{N}(4|\alpha|^2),
\end{split} 
\end{align}
where $L_{n}(x)$ is the $n^{th}$ Laguerre polynomial. 

\section{Entangled photon-added squeezed vacuum states}\label{squeezed_appendix}
When the initial state is chosen to be:
\begin{equation}
\ket{\psi_{in}} = \frac{1}{\sqrt{2}} \left( \ket{\Psi_{+}^{r}}\ket{\Psi_{-}^{r}} + \ket{\Psi_{-}^{r}} \ket{\Psi_{+}^{r}} \right)
\end{equation}
as defined in equation \eqref{squeezed_sec}, the Wigner function of the post-measurement state is give by \eqref{wprod}, with
\begin{align}
\begin{split}
W_{ee}(\beta_r, \beta_i) = \frac{2}{\pi (2 + \tanh^2r)} \exp\left(- 2 \beta_{r}^2e^{2r} - 2 \beta_{i}^{2}e^{-2r} \right) \left\lbrace 16\beta_{r}^{4}e^{4r} + 16\beta_i^{4}e^{-4r} + 32\beta_{r}^{2}\beta_{i}^2 +  \beta_r^2 (-24e^{2r}  + 12\sech r e^{r}  \right. \\ \left. - 8 e^{2r}  + 4e^{3r}\sech r ) +  \beta_i^2 (-24e^{-2r} + 12\sech r e^{-r} - 8 e^{-2r} + 4e^{-3r}\sech r ) + 3 - \sech^2r \right\rbrace ,
\end{split} \\
\begin{split}
W_{eo}(\beta) = \frac{2}{\pi\sqrt{2 + \tanh^2r}} \exp\left(- 2 \beta_{r}^2e^{2r} - 2 \beta_{i}^{2}e^{-2r} \right)  \left\lbrace  8\beta_r^{3}e^{3r} + 8i\beta_i^{3}e^{-3r} + 8i\beta_{r}^{2}\beta_{i}e^{r} + 8\beta_{i}^2\beta_{r}e^{-r} \right. \\ \left. + \beta_{r}(3\sech r - 8e^{r} + e^{2r}\sech r) + i\beta_{i}(3\sech r - 8e^{-r} + e^{-2r}\sech r)   \right\rbrace,
\end{split} \\
\begin{split}
W_{oe}(\beta_{r}, \beta_i) = W_{eo}^{*}(\beta_r, \beta_i),
\end{split} \\
\begin{split}
W_{oo}(\beta_r, \beta_i) =  \frac{2}{\pi} \exp\left(- 2 \beta_{r}^2e^{2r} - 2 \beta_{i}^{2}e^{-2r} \right) \left\lbrace 4\beta_{r}^{2}e^{2r} + 4\beta_i^{2}e^{-2r} - 1 \right\rbrace .
\end{split} 
\end{align}

\end{document}